\documentclass[aps,floatfix,prd,%showpacs,preprintnumbers,twelvepoint
 ]{revtex4}
\usepackage{graphicx}
\usepackage{color}
\usepackage{slashed}
\usepackage{amsmath}
\usepackage{multirow}

\newcommand\bef{\begin{figure}}
\newcommand\eef[1]{\label{fg:#1}\end{figure}}
\newcommand\beq{\begin{equation}}
\newcommand\eeq[1]{\label{#1}\end{equation}}
\newcommand\beqa{\begin{eqnarray}}
\newcommand\eeqa[1]{\label{#1}\end{eqnarray}}
\newcommand\bet{\begin{table}}
\newcommand\eet[1]{\label{tb:#1}\end{table}}

\newcommand\fgn[1]{Figure \ref{fg:#1}}
\newcommand\eqn[1]{eq.\ (\ref{#1})}
\newcommand\scn[1]{Section \ref{sec:#1}}
\newcommand\apx[1]{Appendix \ref{apx:#1}}
\newcommand\tbn[1]{Table \ref{tb:#1}}

\newcommand\A{{\cal A}}
\newcommand\B{{\scriptscriptstyle B}}
\newcommand\bilin[1]{\overline\psi{#1}\psi}

\newcommand\I{{\cal I}}
\newcommand\ie{{\sl i.e.\/}}
\newcommand\K{{\scriptscriptstyle K}}

\newcommand\Lhf{L_{\scriptscriptstyle{HF}}}
\newcommand\Lgb{L_{\scriptscriptstyle{pGB}}}

\newcommand\N{{\cal N}}

\newcommand\ppbar{\langle\overline\psi\psi\rangle}
\newcommand\tco{T_{co}}
\newcommand\Tr{{\rm Tr}\/}
\newcommand\U{{\cal U}}

\begin{document}

\title{An effective field theory for thermal QCD 
 with 2+1 flavours}
\author{Sourendu\ \surname{Gupta}}
\email{sgupta@theory.tifr.res.in}
\affiliation{International Center for Theoretical Sciences,\\ Tata Institute
	of Fundamental Research,\\ Survey 151 Shivakote, Hesaraghatta Hobli, 
	Bengaluru North 560089, India.}
\author{Pritam \surname{Sen}}
\email{spsps3333@iacs.res.in}
\affiliation{School of Physical Sciences, Indian Association for the
         Cultivation of Science,\\ 2A and 2B Raja S.\ C.\ Mullick Road,
	 Kolkata 700032, India.}
\author{Rishi\ \surname{Sharma}}
\email{rishi@theory.tifr.res.in}
\affiliation{Department of Theoretical Physics, Tata Institute of Fundamental
         Research,\\ Homi Bhabha Road, Mumbai 400005, India.}
\begin{abstract}
We write a long-distance effective field theory (EFT) for QCD at finite
temperature just below the crossover temperature $T_c$.  The low energy
constants (LECs) of this EFT are obtained from lattice measurements of the
screening mass of pions at two temperatures for $N_f=2+1$ using lattice
results obtained at physical values of pion and Kaon masses,
and $N_f=2$ where the lattice simulations were performed
with a heavier pion mass.  The EFT gives good predictions for other
static pion properties for $N_f=2$, where lattice results are available.
We show the corresponding predictions for $N_f=2+1$, where they are not
yet measured.  We demonstrate that EFT gives excellent predictions for
the phase diagram in $N_f=2+1$.  The predictions for the pressure are
investigated, and predictions are also given for a Wick-rotated real-time
quantity called the kinetic mass.
\end{abstract}
\maketitle

\goodbreak\section{The Effective Field Theory}\label{sec:one}

Extremely detailed results for thermal QCD are now available from lattice
computations at finite temperature, $T$. However there are parts of the
phase diagram of QCD which remain outside the reach of direct lattice
computations.  Among the outstanding problems is to compute directly
phase diagram at finite (real) baryon chemical potential. The same sign
problem which arises in this case also arises when trying to compute
the phase diagram at finite isospin chemical potential when the light
quarks are allowed to take different masses. A much bigger sign problem
arises in trying to compute the real-time dynamics of thermal QCD on
the lattice. This analytic continuation from Euclidean to Minkowski
metric promises to provide the answer to many questions of dynamics
near equilibrium. While we are unable to answer all the questions that
a complete method would permit, we explore one avenue of systematic
expansions. This is to use a low-energy effective field theory (EFT)
to capture accurately the physics below an UV cutoff $\Lambda$. The
effects of the UV modes are captured in the low energy constants (LECs)
which appear in the Lagrangian of the EFT. The LECs are tuned using
lattice computations at finite temperature, and the EFT Lagrangian is
then used to extract physics below the scale $\Lambda$ in domains where
lattice methods are unavailable.

The key to using EFTs is to be able to identify central features of
the physics which can be easily captured. For an EFT in the presence of
matter this is the observation that Lorentz invariance has to be given
up because there exists a special frame in which the center of mass
of matter is at rest \cite{Weldon:1982aq}. A relativistic theory will
remain Lorentz covariant, and the important issue of the counting of
mass dimensions of operators will be the same as in a theory in vacuum.
Two key physics issues are easily captured in such a formulation. First,
that the difference between a pole and a screening mass are captured
through a low-energy constant (LEC). Second, if we are interested in
a gauge theory, such as QED in matter, then gauge invariance can allow
longitudinal polarization, and hence change the polarization sums in
loops. The EFTs that we write will retain the full rotational symmetry
and the discrete groups CPT.

We write bottom-up EFTs for QCD at finite temperature which try to
capture the chiral symmetry breaking and restoration involved in its
phase diagram. The chiral symmetry group of relevance is determined by
the texture zeroes of the quark mass matrix. 
With $N_f$ flavours of chiral quarks, the global symmetry of QCD is
$U_B(1)\times SU_L(N_f)\times SU_R(N_f)$. An effective field theory
which realizes all these symmetries below an UV cutoff $\Lambda$ can be
written in terms of effective quark field spinors $\psi$ with 4 Dirac
components, $N_f$ flavour components, and $N_c$ colour components,
resulting in a net dimension $\N=4N_fN_c$. In lukewarm QCD, \ie, for $T$
close to and largely below $T_c$, it has been argued that the effect of
gluons may be neglected, so that there is no colour dynamics in the EFT.
Nevertheless, we carry the $N_c$ components of $\psi$ to allow comparison
with the large-$N_c$ counting which has been used in this region. The
use of quark fields allows us to couple the EFT to chemical potentials,
and thereby extend our work to other parts of the phase diagram.

In order to build the two-flavour EFT we wrote first the mass and kinetic
terms
\beq
 L_3 = d_3\Lambda\bilin{}, \qquad{\rm and}\qquad
 L_4 = \bilin{\slashed\partial_4} + d_4\bilin{\slashed\nabla},
\eeq{free}
where ${\bf m}=d_3\Lambda\I$ is the mass matrix for two degenerate quarks,
each of mass $m_0=d_3\Lambda$. Through this paper we shall use the indices
$i=1$, 2, and 3 for the components of spatial vectors and the index 4 for
the Euclidean time. The notation used in $L_4$ is $\slashed \partial_4
= \gamma_4 \partial_4$ and $\slashed \nabla = \gamma_i \partial_i$,
where repeated dummy indices are summed. We follow the conventions
of \cite{wein, Gupta:2017gbs}. The first term in $L_4$ would define
the normalization of the quark field in any future top-down attempt to
derive the EFT from QCD. The appearance of the low energy constant (LEC)
$d_4$ in $L_4$ is the origin of the difference between a screening mass
and a pole mass. This important aspect of thermal physics arises from
the breaking of boost invariance. The subscripts on the pieces of the
Lagrangian $L$ denote the mass dimension, $D$, of the operators. For $D=3$
and 4, these terms exhaust all the operators allowed by the symmetries.

There are no terms allowed by the symmetries for $D=5$. For $D=6$ there are
two kinds of terms--- $L_6^0$ and $L_6^3$ where the superscript counts the
number of derivatives in the operators. For $N_f=2$ we have
\beqa
\nonumber
 L_6^0 &=&
  \frac{d_{6,1}}{\Lambda^2}\left[ \left(\bilin{}\right)^2 + 
  \left(\bilin{\tau^a(i\gamma_5)}\right)^2\right] +
  \frac{d_{6,2}}{\Lambda^2}\left[\left(\bilin{(i\gamma_5)}\right)^2 +
  \left(\bilin{\tau^a}\right)^2 \right]+\\
\nonumber &&
  \frac{d_{6,3}}{\Lambda^2}\left(\bilin{\gamma_4}\right)^2+
  \frac{d_{6,4}}{\Lambda^2}\left(\bilin{(i\gamma_i)}\right)^2+
  \frac{d_{6,5}}{\Lambda^2}\left(\bilin{\gamma_4\gamma_5}\right)^2+
  \frac{d_{6,6}}{\Lambda^2}\left(\bilin{(i\gamma_i\gamma_5)}\right)^2+\\
\nonumber &&
  \frac{d_{6,7}}{\Lambda^2}
   \left[\left(\bilin{\tau^a\gamma_4}\right)^2 + 
    \left(\bilin{\tau^a\gamma_4\gamma_5}\right)^2\right] +
  \frac{d_{6,8}}{\Lambda^2}
   \left[\left(\bilin{\tau^a(i\gamma_i)}\right)^2 +
    \left(\bilin{\tau^a(i\gamma_i\gamma_5)}\right)^2\right]+\\
&&
  \frac{d_{6,9}}{\Lambda^2}\left[ \left(\bilin{(iS_{i4})}\right)^2 +
    \left(\bilin{\tau^aS_{ij}}\right)^2\right] +
  \frac{d_{6,10}}{\Lambda^2}\left[
    \left(\bilin{\tau^a(iS_{i4})}\right)^2 +
     \left(\bilin{S_{ij}}\right)^2\right],
\eeqa{dim60}
where $S_{ij}$ and $S_{i4}$ are defined in \cite{Gupta:2017gbs}.
The operators with LEC $d_{6,1}$ appear in the NJL model. The rest of
the operators are all allowed by the symmetries of the problem. This is
one of the drawbacks of building bottom-up EFTs: there is a proliferation
of terms and LECs which have to be tamed by other means. The other piece
of the $D=6$ Lagrangian is
\beq
 L_6^3 = \frac{d_{6,11}}{\Lambda^2}\,\bilin{\nabla^2\slashed\nabla}.
\eeq{dim63}
All other terms with three derivatives can be reduced to this using the
equations of motion or eliminated by the symmetries.
The Lagrangian of the EFT up to $D=6$ is $L = L_3 + L_4 + L_6^0 +
L_6^3$. This is a sufficient starting point for $N_f=2$. After spontaneous
symmetry breaking it gives the correct SU(2) vector symmetry from which
a pion EFT can be derived. 

In this paper we discuss the extension to three chiral flavours,
$N_f=3$. Famously, the Lagrangian has an emergent symmetry U$_A$(1), so
that the Goldstone bosons are a nonet of pseudo-scalars instead of the
octet. It is well known in the NJL model that the UV symmetry is obtained
when the 't Hooft determinant term is added (see for example the review
of \cite{Klevansky:1992qe}).  Since this term has $D=9$, in the EFT
approach this means that one has to include all the terms allowed up to
$D=9$. In the next section, we present details.  This has a possibility
of a first order phase transition.  We postpone an account of this to
a follow up paper.  When the texture $N_f=3$ is broken to $N_f=2$, the
theory is called $N_f=2+1$, although at sufficiently deep infrared (IR)
it is clearly an $N_f=2$ EFT. In the next section we give an account of
the reduction to a pion theory.

\goodbreak\section{The EFT with strange quarks}\label{sec:two}

In extending the quark EFT to $N_f=3$, the first change is in the
replacement of the flavour SU(2) generators $\tau^a$ by the flavour
generators $T^a$ (with $1\le a\le8$).  The normalization $\Tr(T^a)^2=2$
allows us to identify the Gell-Mann matrices $\lambda^a$ with the
$T^a$. The remaining generator of U(3) is $T^0=(\sqrt{2/3}){\bf 1}$.
$L_3$ term in the EFT changes to include three degenerate quarks. The
form of the $L_4$ and $L_6^3$ terms remain unchanged. The most general
$L_6^0$ terms have the form of \eqn{dim60} with $d_{6,1}=d_{6,2}$ and
$d_{6,9}=d_{6,10}$. This is the origin of an emergent U$_A$(1) symmetry.

In order to make contact with QCD one has to then add higher dimensional
terms in the EFT until this extra symmetry is removed. For $D=7$ there are
no terms which respect the symmetries. For $D=8$ we find three kinds of
structures. One is bilinear in the quark fields and has five derivatives;
we call this $L_8^5$. One is a product of two quark bilinears, each with
a first derivative operator; we call this $L_8^{11}$. The last one is a
product of two quark bilinears, one without derivatives, the other with
two; this we name $L_8^2$. The number of allowed terms in $L_8^{11}$
and $L_8^2$ are very large, so we do not write them down in detail.
All the operators have the emergent U(3) symmetry. There are exactly two
terms allowed at $D=9$ both of which break the emergent symmetry to SU(3).
They are
\beq
 L_9 = \epsilon_{ff'f''}\epsilon_{gg'g''} (\bilin{^f P_R}^g)\left[
	 \frac{d_{9,1}}{\Lambda^5}
	     (\bilin{^{f'}P_R}^{g'})\,(\bilin{^{f''}P_R}^{g''}) +
	 \frac{d_{9,2}}{\Lambda^5}
	     (\bilin{^{f'}P_R S_{ij}}^{g'})\,(\bilin{^{f''}P_R S_{ij}}^{g''})
	 \right] + (L\leftrightarrow R),
\eeq{dim9}
where $P_R=(1-\gamma_5)/2$ is the projection operator on right handed
quarks. The first term was obtained by 't Hooft \cite{tHooft:1986ooh}
and the second by Sch\"aefer \cite{Schafer:2002ty}. No other terms of
this order are allowed by the symmetries of QCD. Any non-zero values of
$d_{9,i}$ lift the accidental degeneracy.

The Lagrangian $L=L_3+L_4+L_6^3+L_6^0+L_8^5+L_8^{11}+L_8^2+L_9$ can be
treated in a Hartree-Fock approximation. This converts all terms into
quadratics in the Fermion fields once the chiral condensate 
$\langle\overline\psi\psi\rangle = \Lambda^3\sigma$ is introduced. 
In the Hartree-Fock approximation one finds for a quartic term with
two flavour-Dirac matrices $\Theta$ and $\Theta'$,
\beq
 (\bilin{\Theta})\,(\bilin{\Theta'}) \overset{{\scriptscriptstyle HF}}=
  -\Lambda^6\sigma^2\left[(\Tr\Theta)\,(\Tr\Theta')-\Tr\Theta\Theta'\right]
  + \Lambda^3\sigma\left[(\Tr\Theta)\,\bilin{\Theta'}
     + (\Tr\Theta')\,\bilin\Theta-2\bilin{\Theta\Theta'}\right].
\eeq{linear}
Using this, we find
\beq
  \Lhf = -\N \Lambda^4\left(d_6\sigma^2+\frac23d_9\sigma^3\right)
      +m\bilin{}+\bilin{\slashed\partial_4}+d_4\bilin{\slashed\nabla}
      +\frac{d_{6,11}}{\Lambda^2}\bilin{\nabla^2\slashed\nabla}
      +\frac{d_8}{\Lambda^4}\bilin{\nabla^4\slashed\nabla},
\eeq{lhf}
where effective LECs are
\beq
  d_6=\N d_{6,1}-d_{6,3}+3d_{6,4}+d_{6,5}-3d_{6,6}
  \qquad{\rm and}\qquad d_9= (1+N_f)(2d_{9,1}(6+\N)+9d_{9,2}).
\eeq{hflecs}
The LECs $d_4$, $d_{6,11}$ and $d_8$ are the same as in $L_4$, $L_6^3$
and $L_8^5$. Since the vacuum has translational invariance, there are no
terms in $\Lhf$ from $L_8^{11}$ and $L_8^2$.  The terms in $d_6$ and $d_9$
which are linear in $\N$ can be obtained in the Hartree approximation; the
remaining come from exchange (Fock) terms. In terms of the Hartree-Fock
effective LECs, the quark mass
\beq
  m = \left(d_3 + 2d_6\sigma + d_9\sigma^2\right)\Lambda.
\eeq{hfmass}
Since $\Lhf$ is quadratic in quarks, a one-loop evaluation of its free
energy is exact. Since the terms in $d_{6,11}$ and $d_8$ are down by
powers of $\Lambda$, we will use the remaining terms to define quark
propagators, and treat these two terms in a perturbative expansion.
With the free energy we can investigate the self-consistent solutions for
$\sigma$, \ie, the gap equation. We can also find the phase structure of
the theory in this approximation. The cubic term in $\sigma$ certainly
opens up the possibility of a first order phase transition. In a
forthcoming paper we will show that that when $d_9$ is large enough to
push $\eta'$ beyond the UV cutoff $\Lambda$, then it pushes the first
order transition to a region where the pseudo-Goldstone masses are around
an MeV.

\subsection{$N_f=2+1$}

For $N_f=2+1$ it is useful to group the flavour generators into three
sets. We reserve the notation $T^a$ to mean $1\le a\le3$. The notation
$T^m$ will be used with $4\le m\le7$, and the remaining generators
will be always written as $T^8$ and $T^0$. We we need to introduce the
projection operator on the strange quark subspace, $\Pi_s$, and the
complementary operator on the light quarks, $\Pi_\ell=1-\Pi_s$. Note that
$\Pi^\ell T^a\Pi^\ell$ corresponds to $\tau^a$ in the light quark space,
and vanishes in the strange quark space. This is an example of the more
general fact that every generator is either zero or a multiple of identity
in the one-dimensional strange quark space.  Using $\Pi_{\ell,s}$ one
can decompose every quark bilinear into a sum of two terms: one for the
strange quark and the other for light quarks. Clearly, the mass matrix
can be decomposed as
\beq
 \bilin{{\bf m}} = d_3^\ell\Lambda\bilin{_\ell}_\ell 
   + d_3^s\Lambda\bilin{_s}_s.
\eeq{masstexture}
The $D=4$ terms decompose similarly, giving two LECs $d_4^\ell$ and
$d_4^s$.  The same happens in $L_8^3$. However, in the products of
bilinears which enter into $L_6^0$, each LEC $d_{6,i}$ of the $N_f=3$
flavours decomposed into the three LECs $d_{6,i}^{\ell\ell}$ with both
bilinears in the light quark space, $d_{6,i}^{ss}$ with both bilinears
for the strange quark, and $d_{6,i}^{\ell s}$ which is the product of
a light quark operator and a strange quark operator. As an example, the
NJL-model term decomposes as
\beqa
\nonumber
 \frac{d_{6,1}}{\Lambda^2}\left[(\bilin{})^2+(\bilin{(i\gamma_5T^i)})^2\right] 
 &\longrightarrow&
       \frac{d_{6,1}^{\ell\ell}}{\Lambda^2}\left[(\bilin{_\ell}_\ell)^2
              +(\bilin{_\ell(i\gamma_5\tau^a)}_\ell)^2\right] 
       + \frac{d_{6,1}^{ss}}{\Lambda^2}\left[(\bilin{_s}_s)^2
	      +\frac43(\bilin{_s(i\gamma_5)}_s)^2\right] \\
 && \qquad\qquad
       +\frac{d_{6,1}^{\ell s}}{\Lambda^2}\left[\bilin{_s}_\ell\;\bilin{_\ell}_s
              + \bilin{_s(i\gamma_5T^m)}_\ell\;\bilin{_\ell(i\gamma_5T^m)}_s\right].
\eeqa{njlterm}
The same kind of structure is found for $L_8^{11}$ and $L_8^2$, but
$L_8^5$ decomposes like $L_6^3$.  The flavour determinants in $L_9$
ensure that there are no multiplicity of LECs $d_{9,i}$.

Next we examine the Hartree-Fock Lagrangian for $N_f=2+1$. Since the
condensate has the same symmetry as the mass term in \eqn{masstexture},
we may write the theory in terms of light and heavy condensates,
$\sigma_\ell$ and $\sigma_s$ respectively. The matrix of condensates
\beq
 \left\langle\overline\psi_a\psi_b\right\rangle = \Lambda^3\Sigma,
 \qquad{\rm where}\qquad \Sigma=\sigma_\ell\Pi_\ell+\sigma_s\Pi_s.
\eeq{matcond}
For $L_6^{\ell\ell}$ the decomposition works as for $N_f=2$
\cite{Gupta:2017gbs}.  Since the strange quark is in a one-dimensional
subspace of the flavour space, the trace over flavour is trivial, and
all traces in \eqn{linear} reduce to Dirac traces.

The coupling between the light and heavy quarks comes only from
$L_6^{\ell s}$.  Since the flavour structure for this can only involve
$T^m$ with $4\le m\le7$,  Then using the flavour projection operators
we can simplify this in the Hartree-Fock approximation to
\beqa
\nonumber
 (\bilin{_\ell T^m\Gamma}_s)\,(\bilin{_sT^m\Gamma}_\ell) 
    &\overset{{\scriptscriptstyle HF}}=& \Lambda^6\sigma_\ell\sigma_s\Tr(\Gamma\Gamma)
      -\Lambda^3\sigma_\ell\bilin{_s\Gamma\Gamma}_s\\
 && \quad-
  \Lambda^3\sigma_s\left[(\delta_{m4}+\delta_{m5})\bilin{_u\Gamma\Gamma}_u
    +(\delta_{m6}+\delta_{m7})\bilin{_d\Gamma\Gamma}_d\right].
\eeqa{contract2p1dir}
where $\Gamma$ is a Dirac matrix, and we have used different spinors for
the u, d, and s flavour components. We can pair the Dirac matrices into
the following sets
\beq
 1 + (i\gamma_5)^2, \qquad 
 \gamma_4^2 + (\gamma_4\gamma_5)^2, \qquad
 (i\gamma_i)^2+(i\gamma_i\gamma_5)^2, \qquad
 (S_{ij})^2 + (iS_{ij}\gamma_5)^2,
\eeq{pairs}
where we have used the relation $S_{k4}=\epsilon_{ijk4}S_{ij}\gamma_5$ in
the last pair. Since $\gamma_5$ anticommutes with all the $\gamma_\mu$,
and $\gamma_5^2=1$, one finds that each pair gives a vanishing
contribution to $\Lhf$. So the mixing terms between the light and heavy
sectors vanish in the MFT because of the emergent symmetry.  The coupling
between the two condensates then comes only through the $D=9$ term.

As a result, the Hartree-Fock Hamiltonian is
\beqa
\nonumber
  \Lhf &=& -\N \Lambda^4\left(d_6^\ell\sigma_\ell^2
      +d_6^s\sigma_s^2+\frac23d_9\sigma_\ell^2\sigma_s\right)
      +m_\ell\bilin{_\ell}_\ell+\bilin{_\ell\slashed\partial_4}_\ell
      +d_4^\ell\bilin{_\ell\slashed\nabla}_\ell
      +\frac{d_{6,11}^\ell}{\Lambda^2}\bilin{_\ell\nabla^2\slashed\nabla}_\ell
      +\frac{d_8^\ell}{\Lambda^4}\bilin{_\ell\nabla^4\slashed\nabla}_\ell\\
    && \qquad
      +m_s\bilin{_s}_s+\bilin{_s\slashed\partial_4}_s
      +d_4^s\bilin{_s\slashed\nabla}_s
      +\frac{d_{6,11}^s}{\Lambda^2}\bilin{_s\nabla^2\slashed\nabla}_s
      +\frac{d_8^s}{\Lambda^4}\bilin{_s\nabla^4\slashed\nabla}_s,
\eeqa{lhf21}
where the definition of $d_6$ in \eqn{hflecs} is replaced by
\beq
  d_6^\ell = \frac23\N d_{6,1}^{\ell\ell} - d_{6,3}^{\ell\ell} 
    +3 d_{6,4}^{\ell\ell} + d_{6,5}^{\ell\ell} -3 d_{6,6}^{\ell\ell},
    \qquad{\rm and}\qquad
  d_6^s = \frac13\N d_{6,1}^{ss} - d_{6,3}^{ss} 
    +3 d_{6,4}^{ss} + d_{6,5}^{ss} -3 d_{6,6}^{ss}.
\eeq{hflecs21}
Finally, the two effective masses are
\beq
  m_\ell = \left(d_3^\ell + 2d_6^\ell\sigma_\ell + d_9\sigma_\ell\sigma_s\right)
      \Lambda,\qquad{\rm and}\qquad
  m_s = \left(d_3^s + 2d_6^s\sigma_s + d_9\sigma_\ell^2\right)\Lambda.
\eeq{hfmass21}
A detailed analysis, which will be presented in a separate paper, shows that
the coupling between the light and strange sectors causes a first order
transition to appear at very small pseudo-Goldstone masses. However, for
values of $d_9$ which push the $\eta'$ mass above the UV cutoff, values of
$d_3^\ell$ and $d_3^s$ relevant to QCD has the same phase structure that was
seen in the $N_f=2$ theory \cite{Gupta:2017gbs}.

\subsection{The EFT of pseudo-Goldstone bosons}

After chiral symmetry breaking one can introduce small fluctuations about
the condensates through
\beq
 \psi\to\U\psi, \qquad\overline\psi\to\overline\psi\U^\dag \quad
 {\rm where}\quad\U = \exp[i\gamma_5T^i\phi_i/(2f_i)],
\eeq{defgb}
$T^i$ are the generators of the remaining vector flavour symmetry,
and we have allowed for the possibility that the decay constants $f_i$
have different values for $f_a$, $f_m$, $f_8$ and $f_0$ in the case of
$N_f=2+1$. Instead of using $\U$, one can project on the left and right
spinors and use the transformations
\beq
 \psi_L\to U\psi_L, \qquad\psi_R\to U^\dag\psi_R \quad
 {\rm where}\quad U = \exp[iT^i\phi_i/(2f_i)].
\eeq{defgbfl}
The transformation matrix $\U$ is non-trivial in Dirac space, whereas the
matrix $U$ is only a flavour transformation and is trivial in Dirac space.
The partition function for the group-valued field $U$ uses its Haar measure.

Using $U$ in $L$, along with the solutions of the gap equation obtained
using $\Lhf$ gives a Lagrangian which couples the $\phi_i$ to the
quark fields. However, this over counts the degrees of freedom, since
the $\phi_i$ are just a parametrization of the most easily excited
fluctuations in the quark fields. So one needs to integrate over the
quarks in order to reach the target, $\Lgb$, which is the action for
the pseudo-Goldstone bosons. We find that it has the form
\beqa
\nonumber
 \Lgb &=& \frac12\Lambda^2\left[c_2^a\phi_a^2 + c_2^m\phi_m^2
  + c_2^8\phi_8^2 + c_2^0\phi_0^2\right] + \frac12\left[
    \dot\phi_a^2 + \dot\phi_m^2 + \dot\phi_8^2 + \dot\phi_0^2\right]\\
\nonumber && \qquad
  + \frac12\left[c_4^a(\nabla\phi_a)^2 + c_4^m(\nabla\phi_m)^2
    + c_4^8(\nabla\phi_8)^2 + c_4^0(\nabla\phi_0)^2\right]
  + \frac18\left[c_{41}^a\phi_a^4 + c_{41}^m\phi_m^4
    + c_{41}^8\phi_8^4 + c_{41}^0\phi_0^4\right] \\
&& \qquad\qquad
  + \frac14\left[c_{41}^{am}\phi_a^2\phi_m^2 
    + c_{41}^{a8}\phi_a^2\phi_8^2 + c_{41}^{a0}\phi_a^2\phi_0^2
    + c_{41}^{m8}\phi_m^2\phi_8^2 + c_{41}^{m0}\phi_m^2\phi_0^2
    + c_{41}^{80}\phi_8^2\phi_0^2 \right] + \cdots
\eeqa{lgb}
The quantum numbers of $\phi_0$ and $\phi_8$ allow mixing, but we will
show in the next subsection that the $\eta'$ can be decoupled easily and
this mixing will not play a role.
Integrating out the quarks to one-loop order one can write expressions
for the LECs of the pseudo-Goldstone bosons in terms of those for the
quark. In addition, by requiring the normalization of the time derivative
terms to be as shown, one obtains the constants $f_i$ in \eqn{defgbfl}, by
a natural extension of the argument in \cite{Gupta:2017gbs}. Since the LEC
$d_3^s$ is not small, one cannot apply chiral power counting to $\Lgb$.
We have organized it in the mass dimension $D$, and written all the terms
up to $D=4$. By matching a sufficient number of these LECs to measurements
from the lattice, one can derive the LECs of the quark theory.

\subsubsection{The kinetic terms for pseudo-Goldstone bosons}

We consider first the contributions to the kinetic terms in $\Lgb$. These
can arise from $L_4$, $L_8^{11}$ and $L_8^2$.  To begin with, note that
$UU^\dag=1$ implies that the combination $U(i\partial_\mu)U^\dag$ is
Hermitean in flavour space, and an expansion shows that it reduces to
$\partial_\mu\phi_i$ on expanding the exponential. Contraction of the
quark field operators in $L_4$ then gives for each field $\phi_i$
\beq
  \Lgb(4) = -\frac{\Lambda^2}{f_i^2}
      \left[\I_4(\dot\phi_i)^2 + d_4^2\I_3 (\nabla\phi_i)^2\right]
     = \frac12(\dot\phi_i)^2 + \frac12c_4^i(\nabla\phi_i)^2,
\eeq{kineticterms}
where we have suppressed the light and strange quark identifiers in $d_4$
and the integrals $\I_3$ and $\I_4$ (which are given in \apx{four}).
With all this, one finds simply
\beq
 \frac{f_a^2}{\Lambda^2} = -2\I_4^{\ell\ell}, \;\;
 c_4^a = (d_4^\ell)^2\;\frac{\I_3^{\ell\ell}}{\I_4^{\ell\ell}}, \quad
 \frac{f_m^2}{\Lambda^2} = -2\I_4^{\ell s}, \;\;
 c_4^m = d_4^\ell d_4^s\;\frac{\I_3^{\ell s}}{\I_4^{\ell s}}, \quad
 \frac{f_8^2}{\Lambda^2} = -\frac23(\I_4^{\ell\ell}+2\I_4^{ss}), \;\;
 c_4^8 = \frac{(d_4^s)^2\I_3^{\ell\ell}+2(d_4^s)^2\I_3^{ss}}{\I_4^{\ell\ell}+2\I_4^{ss}}.
\eeq{scales}
For the light quarks this reproduces the results of \cite{Gupta:2017gbs}.

$L_8^{11}$ and $L_8^2$ can clearly give additional contributions to
$c_4^a$, $c_4^m$ and $c_4^8$ through the same mechanism. However, there
are four quark fields to be contracted, so there are no contributions
at one-loop order.  The two and three loop integrals are complicated,
but by dimensional arguments it can be shown that they are down by a
power of $(T/ \Lambda)^4\times N_c/(4\pi^2)$ for each added loop order.
The terms $L_6^3$ and $L_8^5$ contain higher derivative terms but
only two quark fields. They give contributions to these LECs, but the
extra derivatives act on the quarks and give more powers of momentum
in the loops, and are therefore suppressed by $(T/\Lambda)^2$ and
$(T/\Lambda)^4$ respectively.

\subsubsection{Pseudo-Goldstone Boson masses}

\bef
\begin{center}
\includegraphics[scale=0.7]{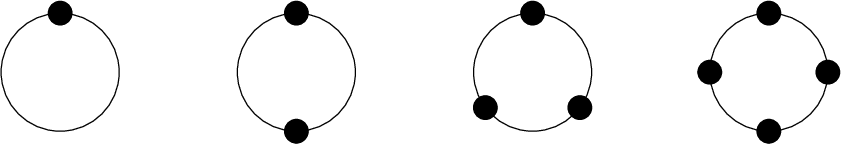}
\end{center}
\caption{The first four terms organized by the number of insertions of
$\bf M$ into the one-loop expression for $\Lgb(3)$. Each of these can
be expanded in powers of the $\phi^i$s. If we retain only terms up to
the fourth power, then these are the only insertions which need to be
considered.}
\eef{massins}

We have neglected the LECs for the $\eta'$ in the previous sub-section,
in anticipation of its decoupling.  In this section we will show
how it does so.  Formally the LECs $c_2$ get contributions from $L_3$
and $L_9$. Introducing fluctuations through \eqn{defgbfl} gives the
contribution of $L_3$ as
\beq
 \Lgb(3) = \bilin{{\bf M}}, \qquad{\rm where}\qquad 
   {\bf M}=(U{\bf m}U-{\bf m})P_L + (U^\dag{\bf m}U^\dag-{\bf m})P_R,
\eeq{massins}
where $P_{L,R}$ are the left and right helicity projectors for the quarks.
To one loop order one can organize this by the number of insertions of
$\bf M$ into the quark loop, as shown in \fgn{massins}. Note that every
insertion of $\bf M$ can be decomposed by writing it as $(\Pi_\ell
+ \Pi_s){\bf M}(\Pi_\ell + \Pi_s)$, so each of the topologies in
\fgn{massins} gives rise to diagrams with zero, one, or two strange
quarks. A straightforward computation then gives
\beq
 c_2^a=\frac23\N\left(\frac\Lambda{f_a}\right)^2\left[d_3^\ell\sigma_\ell
  -(d_3^\ell)^2\I_1^{\ell\ell}\right],
\eeq{pionmass}
where the integral $\I_1^{\ell\ell}$ is given in \apx{four}. Since it is
regular in the limit $d_3^\ell\to0$, the pion mass vanishes in the chiral
limit, and a thermal version of the Gell-Mann-Oakes-Renner (GMOR) relation
is obtained from the leading term in \eqn{pionmass}. By systematically
taking other flavour projections in $\bf M$ one similarly obtains $c_2^m$,
$c_2^8$ and $c_2^0$. It is interesting to take the $N_f=2$ chiral limit
by sending $d_3^\ell\to0$ while holding $d_3^s$ fixed. In this limit
$c_2^m$, $c_2^8$ and $c_2^0$ are finite. Additionally, if $d_3^s$ is
small, then they are linear in $d_3^s$.

Clearly powers of $\phi_i$ in $\Lgb$ can only come from $L_3$, $L_6^0$,
and $L_9$ terms in the quark EFT, since the other terms all involve
derivatives.  $L_6$ is fully invariant under the symmetries, and hence
gives no contributions in $U$.  Since $L_9$ is invariant under SU(3) but
not under the overall U(1) phase, it has a non-vanishing contribution
which can be expanded in powers of $\phi_0$.  Then using the fact that
the expansion is made around the solution of the gap equation obtained through
$\Lhf$, and expanding to quadratic order in $\phi_0$, one finds only an
additional contribution to $c_2^0$ from
\beq
 \Lgb(9) = -\N d_9\frac{\Lambda^4}{f_0^2}
     \sigma_\ell^2\sigma_s\phi_0^2.
\eeq{etapmass}
This is a pleasant result, since it shows that the mass of the undesired
field $\phi_0$ may be pushed above the UV cutoff $\Lambda$ by tuning
$d_9$, without changing the rest of $\Lgb$ since $d_9$ only appears
explicitly here. With this, the mixing of $\phi_0$ and $\phi_8$
is also removed from the EFT, and the latter becomes the pure $\eta$ meson
state. The effect of $d_9$ continue to be felt in $\Lgb$ since the
solutions of the gap of equations, namely the values of $\sigma_\ell$
and $\sigma_s$ depend on $d_9$. Since the $\eta'$ mode can be decoupled
easily, we do not consider it in the rest of this discussion.

\subsubsection{The coupling terms for pseudo-Goldstone bosons}

The LECs $c_{41}$, $c_{41}^m$, {\sl etc\/}, come from the expansion of
the exponentials in $\bf M$ as explained earlier. In \cite{Gupta:2017gbs}
it was shown that for $N_f=2$ one-loop contributions to $c_{41}$ have
pieces which scale with different powers of $d_3^\ell$, ranging from one
to four. These come from the topologies shown in \fgn{massins}. In the
$N_f=2+1$ theory, the flavour projections are an only extra complication.
Handling them is tedious but does not require new techniques. For the pion
self-coupling, the result is
\beq
  c_{41}^a = -\frac{m_\pi^2}{3f_a^2}+\frac23\left(\frac\Lambda{f_a}\right)^4
    (2d_3^\ell)^2\left(4\I_1^{\ell\ell}+3\I_2^{\ell\ell}\right) +\cdots,
\eeq{coupling}
where the integrals $\I_1$ and $\I_2$ are discussed in \apx{four},
and we have written down the results from the first two topologies of
\fgn{massins}.  The remaining diagrams give contributions of higher
order in $d_3^\ell$. For the other couplings the propagators change
due to the flavour projections at the vertices; leading to changes
in these two integrals (see \apx{four}). The IR and UV properties of
the integrals are unchanged. The expansions are more generally joint
expansions in $d_3^\ell$ and $d_3^s$, but the sum of the powers of the two
do not exceed 4. In the light quark chiral limit, when $d_3^\ell\to0$,
holding $d_3^s$ finite, $c_{41}^a$ and $c_{41}^{a8}$ vanish, but none
of the others do. It is interesting to note that in this limit \beq
  c_{41}^m=-\frac{c_2^m\Lambda^2}{3f_m^2}\qquad{\rm and}\qquad
  c_{41}^8=-\frac43\,\frac{c_2^8\Lambda^2}{9f_8^2}.
\eeq{coups} Also, in the same limit, $c_{41}^{am}\propto d_3^s$, and
close to $\tco$ becomes $-m_\K^2/(12 f_a^2)$.

The scale factors and all the LECs of the pseudo-Goldstone bosons are
directly computable in thermal QCD, and so can be used to match $\Lgb$
to lattice computations. Furthermore, the integral expressions here can
then be used to match them to the LECs of the quark EFT.

\subsubsection{The pion EFT}

\bef
\begin{center}
 \includegraphics[scale=0.65]{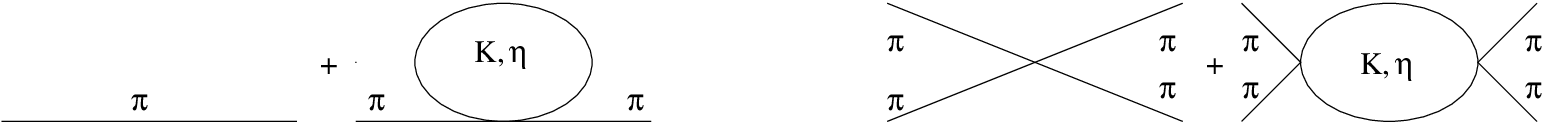}
\end{center}
\caption{The pion EFT is obtained by integrating over all hard modes
 in an energy shell between $\Lambda_{2+1}$ and $\Lambda_2$. The main
 constraints on the former is that it must lie between the proton and
 $\eta$ masses. On the other hand, $\Lambda_2$ lies below the Kaon mass
 and must be larger than $\tco$ so that it catches thermal physics
 in this range. The main corrections to the pion 2-point and 4-point
 functions are shown. The one loop correction is resummed using a
 Dyson-Schwinger formulation.}
\eef{hardmodes}

A further simplification is now possible. One can integrate over the
strange mesons and the hard modes of the pions, and so get an effective
pion theory at even smaller energy
\beq
 L_\pi = \frac12\Lambda^2c_2\phi_a^2 + \frac12\dot\phi_a^2 
  + \frac12c_4(\nabla\phi_a)^2 + \frac18c_{41}\phi_a^4 + \cdots
\eeq{lpi}
Thermal effects in the integration of Kaons and $\eta$ are expected to
be exponentially small, and the methods of \cite{Gasser:1984gg} may be
used to accomplish this. However, when the pion mass is realistic, it
is smaller than $\tco$, and one expects chiral power counting to work
in $L_\pi$. In this counting every power of $m_\pi$ scales in the same
way as a derivative, so that the mass and kinetic terms are all of the
same order (leading order, LO). The $c_{41}$s in $\Lgb$ are only one of
several new types of LECs which are obtained at the next-to-leading order
(NLO) in this counting.  Obtaining a consistent power counting again
in the reduced theory would need to include all the other NLO terms
while doing the one-loop integrations. We do not perform this higher
order computation, whose only purpose would be to allow us to express
the LECs of \eqn{lpi} to those of \eqn{lgb}, extended fully to NLO.

Instead, only a few simple facts are needed from knowing this can be
done. First, that the effect of the strange quarks is implicit in the
LECs of $L_\pi$, although strangeness is not explicit in this low-energy
EFT. A second useful point is that the UV cutoff of this EFT, $\Lambda_2$,
is lower than $\Lambda_{2+1}$ which would be appropriate for \eqn{lgb}.
Finally, recall again that in chiral power counting the first three
terms are of LO, whereas the last term is the first of several NLO terms.

\subsection{UV insensitivity of the low-energy theory}

Since the low-energy EFT of \eqn{lpi} which is obtained for $N_f=2+1$
is the same as that obtained in \cite{Gupta:2017gbs} for $N_f=2$,
the low-energy EFT is insensitive to the UV theory. Also, QCD with
$N_f=2+1$ has a crossover at finite temperature, just as QCD with $N_f=2$
does. Then it becomes convenient to treat \eqn{lpi} as if it descending
from an $N_f=2$ quark EFT, since this has a smaller number of LECs. Of
course these LECs will be matched to pion properties in the $N_f=2+1$
lattice computations, so they have implicit knowledge of the effect
of the strange quark on low-energy dynamics. The bonus is that this UV
insensitivity can be utilized by computing the phase diagram of $N_f=2+1$
QCD from this $N_f=2$ quark EFT. In this flavour-reduced quark EFT,
one needs to take into account only the $D=3$, 4, and 6 terms, since the
correct symmetry of the continuum theory is already recovered with $D=6$.
In this section and later, whenever the LECs of the quark EFT are written
without superscripts $\ell$ and $s$, they refer to the flavour-reduced EFT.

Since $\Lhf$ is quadratic in fields, the free energy can be evaluated
exactly in this approximation, and turns out to be
\beqa
\nonumber
 \Omega &=& -\N V\left[ \Lambda^4d_6\sigma^2
   + \frac{m^4}{64\pi^2d_4^3}\left\{
	   \log\left(\frac{m^2}{M^2}\right)-\frac32\right\}
   + \frac{5m^6d_{6,11}}{128\pi^2d_4^6\Lambda^2}\left\{
	   \frac{23}{30}-\log\left(\frac{m^2}{M^2}\right)\right\}\right.\\
   && \qquad \left.+ \frac T{2\pi^2d_4^3}\int_0^\infty 
           dp\,p^2\log\left(1+{\rm e}^{-E_p/T}\right)
   + \frac{d_{6,11}}{2\pi^2d_4^6\Lambda^2}\int_0^\infty 
	   \frac{dp}{E_p}\;\frac{p^6}{1+{\rm e}^{E_p/T}}\right],
\eeqa{enef}
where $E_p^2=m^2+p^2$. The factors of $d_4^3$ in the kinetic term
have been absorbed by the redefinition $p\to d_4p$, and gives rise
to the powers of $d_4$ in front of the integrals. Here $d_{6,11}$ has
been included to linear order. Since the corresponding operator is a
correction to the kinetic term, when taken to all orders, it changes
the definition of $E_p$ and gives
\beq
  E_p^2 = m^2 + p^2 \left(d_4-d_{6,11}\frac{p^2}{\Lambda^2}\right)^2.
\eeq{dispersion}
When $T\ll\Lambda$, then the thermal integrals cut off the
range of momentum which are important to the problem and imply
that $p\ll\Lambda$. Then clearly it is sufficient to expand the
result to leading order in $d_{6,11}$ in order to get \eqn{enef}.
The computation can be easily extended to finite baryon chemical
potential, $\mu_\B$, by recalling that this results in adding the term
$(\mu_\B/N_c)\bilin{\gamma_4}$ to $\Lhf$ for the chemical potential on
the quarks.

\subsection{The phase diagram}

With this, the gap equation can be written down. There is a critical
point only for $d_3=0$. The equation for $T_c$ is obtained by requiring
the second derivative of the free energy with respect to the condensate
$\Sigma$ to vanish.  The integrals over the Fermi distribution can be
easily performed in this limit. Using the notation $z=\mu_\B^2/\Lambda^2$,
and $t=T_c(\mu_\B)/\Lambda$ one then obtains
\beq
  \frac{\pi^2d_{6,11}}{d_4^3}\,\left(\frac72t^4+\frac5{3\pi^2}\,zt^2
     +\frac5{54\pi^4}\,z^2\right) + \left(t^2+\frac z{3\pi^2}\right)-t_0^2 =0
  \qquad{\rm where}\qquad t_0^2=\frac{12d_4^3}{d_6}. 
\eeq{critcurv}
In \cite{Gupta:2017gbs} we had considered the case with
$d_{6,11}=0$. Retaining only the positive solution of the quadratic in
this limit, one has $T_c=t_0\Lambda$. In this limit there is a line of
second order transitions,
\beq
  \left(\frac{T_c(\mu_\B)}{T_c}\right)^2 = 1 - 
      \frac3{N_c^2\pi^2}\, \left(\frac{\mu_\B}{T_c}\right)^2,
\eeq{ellipse}
where we continue to use the lighter notation $T_c$ for $T_c(\mu_\B=0)$.
Since this is the equation of an ellipse in the phase diagram of $T$
versus $\mu_\B$, we call this the chiral critical ellipse. This is the
phase diagram of a generic NJL model, \ie, a model which has the kinetic
terms and the D=6 four-Fermi terms constrained by the flavour symmetry.

When $d_{6,11}$ is non-vanishing then the gap equation is a disguised
quadratic equation in $t^2$. Negative or complex solutions are discarded,
but there may still be two positive solutions if the discriminant of the
quadratic is larger than $d_4^3$. However, for the EFT to model QCD, there
should be only one acceptable solution for $t$, and this should lie in the
range $0<t<1$. We find that for $d_{6,11} \ge -d_4^3/(14\pi^2t_0^2)$ there
are two positive solutions for $T_c$. At exactly this critical value,
the two solutions are degenerate and give $t=\sqrt2t_0$. With
increasing $d_{6,11}$ one increases and the other decreases towards
zero. There is exactly one positive solution when
\beq
  d_{6,11} > \frac{2d_4^3}{7\pi^2}(t_0^2-1).
\eeq{limitlec}

\bet
\begin{center}\begin{tabular}{|l|l|l|l|}
\hline
Reference            & $\kappa_2$ &  $\kappa_4$ & $\widetilde\kappa_4$ \\
\hline
\cite{Cea:2014xva,Cea:2015cya} & $0.020 (4)$ & & \\
\cite{Bonati:2014rfa,Bonati:2015bha} & $0.0135(20)$ & & \\
\cite{Bonati:2018nut} & $0.0145(25)$ & & \\
\cite{HotQCD:2018pds} $(\partial_T^2\Sigma=0)$
    & $0.015(4)$ & $-0.001(3)$ & $-0.001(3)$\\
$\qquad (\partial_T\chi=0)$ 
	& $0.016(5)$ & $\hphantom{-}0.002(6)$ & $\hphantom{-}0.002(6)$\\
\cite{Borsanyi:2020fev}
	& $0.0153(18)$ & $\hphantom{-}0.00032(67)$ & $\hphantom{-}0.00020(42)$\\
\hline          
\end{tabular}\end{center}
\caption{Recent lattice measurements of the curvature coefficients
(\cite{HotQCD:2018pds} reports results using two methods, as
indicated). The values of $\widetilde\kappa_4$ are derived using
\eqn{tilded} and standard Gaussian error propagation.}
\eet{curvs}

A straightforward computation using the definitions in \eqn{derivs} 
and \eqn{critcurv} then gives us the curvature coefficients
\beq
 \kappa_2 = \frac3{2N_c^2\pi^2}\,\frac{1+5\epsilon}{1+7\epsilon},
 \qquad{\rm and}\qquad
 \widetilde\kappa_4 = -\frac{3\epsilon}{N_c^4\pi^4}\,
    \frac{1+20\epsilon+70\epsilon^2}{(1+7\epsilon)^3}.
\eeq{eftcurvs}
Notice that $\widetilde\kappa_4$ vanishes linearly with $\epsilon =
\pi^2d_{6,11}T_c^2/(d_4^3\Lambda^2)$. It is also interesting to observe
that in the large $N_c$ limit taken together with the chiral limit,
the second order chiral symmetry restoring transition happens at a
$T_c$ which is independent of $\mu_\B$. This is different from the
first order deconfining line found in \cite{McLerran:2007qj}, and it
has been conjectured that it either lies below or is coincident with
it \cite{Datta:2010sq}. Notice, furthermore, that $\widetilde\kappa_4$
is suppressed by two extra powers of $N_c$ compared to $\kappa_2$. As a
result, the chiral critical ellipse may be a good approximation to the
shape of the phase diagram at relatively small $N_c$.

\tbn{curvs} collects the lattice results for the curvature coefficients
which were obtained by different groups using different methods. In
the last decade $\kappa_2$ has begun to converge to a common value,
with the most recent computations being in very good agreement with each
other. The values of $\kappa_4$ are also beginning to be accessible in
lattice measurements, and we quote the currently available values. From
these we extract the values of $\widetilde\kappa_4$. The present data
on this quantity indicates that it is consistent with zero, which is
also consistent with the large $N_c$ power counting. Consistency of
both $\kappa_2$ and $\kappa_4$ as measured on the lattice with the EFT
requires a small value of $d_{6,11}$. In view of this, in the remainder
of this paper we report numerical work with the version of the EFT with
$d_{6,11}$ set to zero. One sees that in this case one has the prediction
\beq
 \kappa_2 = \frac3{2N_c^2\pi^2} \xrightarrow{N_c=3} 0.169,
\eeq{eftpred}
which is consistent with the results of \cite{HotQCD:2018pds,
Borsanyi:2020fev} at the 68\% CL.

\goodbreak\section{Using lattice computations for $N_f=2$}\label{sec:three}

We start by setting out our procedure for determining the LECs
from measurements and then using these in the EFT to produce further
predictions. We choose the number of inputs to be the same as the number
of LECs to be extracted, hence the process amounts to solving three
coupled equations. However, each of the input quantities have errors,
and they propagate to the LECs, and through them to the predictions
of the EFT. So it is numerically easier to treat the extraction as a
fitting process which minimizes $\chi^2$, defined in the usual way
as the sum of the squares of the difference between the theory and
measurement normalized by the measurement error. We check that the
``best fit'' value of $\chi^2$ is the same as the machine precision;
this implies that the input measurements are properly described by the
model. For any other values of the LECs, the value of $\chi^2$ can then be
used as usual to define the 68\%, 95\% and 99\% confidence limits (CLs)
\cite{PressTeukolsky} on the LECs. By a bootstrap sampling within these
CLs, the statistical distribution of EFT predictions can be obtained,
and quoted as CLs on them. All error bars on predictions are obtained
in this way.

Fits to the LECs using what is called the set C1 lattice data for $N_f=2$
\cite{Brandt:2014qqa}, and the predictions which come out of it were given
in \cite{Gupta:2017gbs}.  (The notation $m_\pi$ in \cite{Brandt:2014qqa}
corresponds to our $m_\pi^D$, $u_f$ of \cite{Brandt:2014qqa}
to our $u_\pi$, and the definition of the chiral condensate in
\cite{Brandt:2014qqa} corresponds to $-\N\ppbar$ in our notation.)
The method that we used to extract the LECs in \cite{Gupta:2017gbs},
namely to use one value of $m_\pi^D$ and one of $u_\pi$ as inputs to
the fits cannot be used for $N_f=2+1$, since measurements of $u_\pi$
have not been performed yet for $N_f=2+1$. Here we explore a different
scheme for extracting the LECS.

\subsection{Extracting the LECs by matching lattice data}

\bef
\includegraphics[scale=0.75]{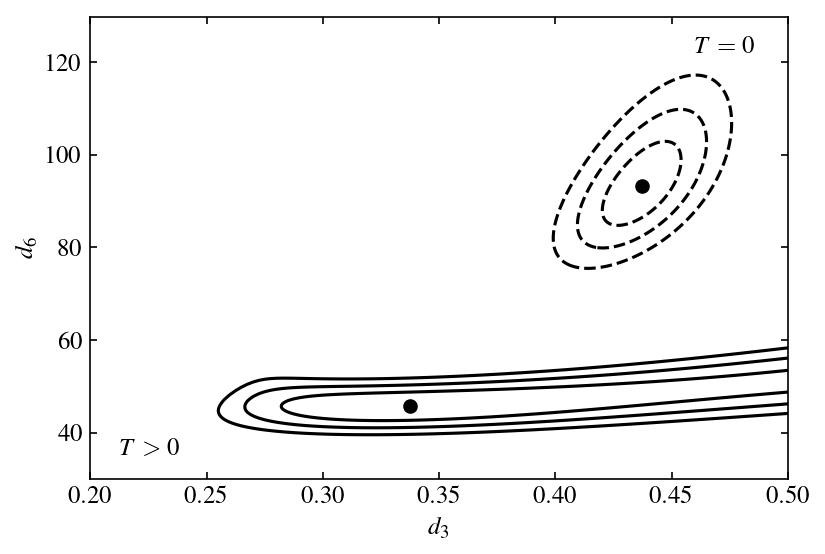}
\caption{We compare extractions of the LECs $d_3$ and $d_6$ at $T=0$
and finite $T$. The best fit values of the LECs are indicated by a dot,
and the successive contours enclose the 68\%, 95\% and 99\% CLs.  For the
finite $T$ theory the best fit value of $d_4=1.21^{+0.09}_{-0.07}$.}
\eef{zerolecs}

The method of extraction of the LECs used in \cite{Gupta:2017gbs}
was geared to the choice $\Lambda=T_c$. However, the value of the
UV cutoff should be flexible, and it is more instructive to take it
to be large enough to include thermal pion effects in full. Here we
will utilize $\Lambda=300$ MeV. Our first extraction of the LECs uses
as inputs the lattice values of $m_\pi^D$ and $u_\pi$ at a $T$ below
$\tco$ and the value of $\tco$.  Another change is that we now use the
Schwinger-Dyson resummed expression of $m_\pi^D$ \cite{Gupta:2020zqo}
for these extractions.  

The EFT can also be adapted to $T=0$ by restoring full Lorentz invariance
(\ie, $d_4=1$ and some of the $d_{6,\A}$ are degenerate) then one has
only two couplings to determine in the corresponding $\Lhf$, namely $d_3$
and $d_6$. They can be determined from the $T=0$ values of $m_\pi$ and
the pion decay constant $f_\pi$. We use as input into the determination
of $T=0$ LECs the lattice data at bare couplings corresponding to those
used in the finite $T$ computations. In \fgn{zerolecs} we compare the
LECs obtained at $T=0$ with those obtained at finite temperature using
the method of \cite{Gupta:2017gbs}. In both these extractions we have
used $\Lambda=300$ MeV.

Note that the best fit range of $d_3$ at $T=0$ and $T>0$ are completely
compatible with each other. In this case the major effect of temperature
is a large shift in $d_6$. We will utilize this observation to extract the
LECs of the finite temperature EFT from lattice data in another way. This
will be done in two stages, first by using the $T=0$ data to extract the
$d_3$ and $d_6$. Next the range of $d_3$ obtained in this way is taken over
to finite temperature where the remaining LECs, namely $d_4$ and $d_6$, are
obtained by fitting to lattice measurements of $m_\pi^D$ at two nearby
value of $T$ below $\tco$. This changes the best fit values of the LECs,
as is to be expected. However, the predictions of physical quantities does
not change much, as we next show.

\subsection{The phase diagram}

\bef[h]
\includegraphics[scale=0.5]{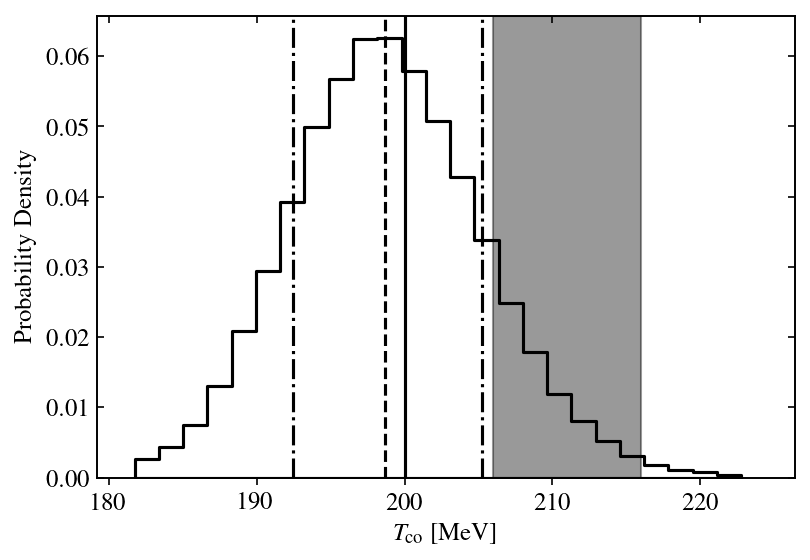}
\includegraphics[scale=0.5]{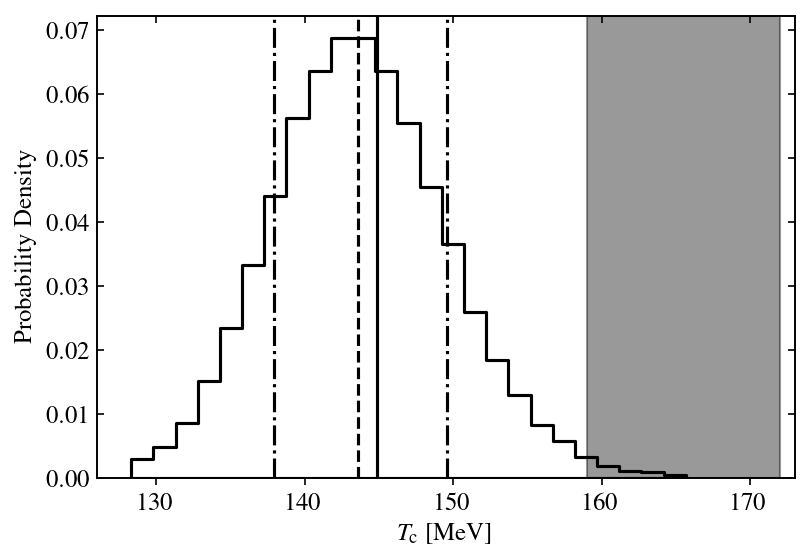}
\caption{Predictions for $\tco$ and $T_c$ using the fitted LECs shown
as histograms obtained by sampling the 90\% CLs of the fits.  The median
value and the limits of the 68\% band for the EFT predictions are shown
with broken vertical lines. The continuous vertical line shows the best
fit value of the LECs, and the gray band is the lattice extraction of
the corresponding quantity \cite{Brandt:2014qqa}.}
\eef{tcopred}

With this scheme only static pion properties go into the determination
of the LECs.  The first prediction that we can get is for $\tco$
(which is the temperature at which the chiral susceptibility peaks),
and its limiting value in the chiral limit, $T_c$. The histograms for
these predictions are shown in \fgn{tcopred}, when points are sampled
within the $\Delta\chi^2$ range of $d_4$ and $d_6$ corresponding to
the 90\% CL with weight proportional to $\Delta\chi^2$. The skewness of
the distributions are seen in two ways. First the upper and lower edges
of the 68\% CLs are not symmetric around the median. Also, due to the
skewness the best-fit LECs give slightly different predictions than the
median. However, these differences are mild.

The predicted value for $T_c$ is now $144\pm6$ MeV, which is somewhat
below the value of 170 MeV reported in \cite{Brandt:2014qqa}.
From \fgn{tcopred} it is clear that they are quite compatible within
95\% CL limit.  However, the value of $T_c$ obtained in the EFT and that
quoted in \cite{Brandt:2013mba} differ significantly.  The latter were
obtained using O(4) exponents. Using mean field exponents instead would
decrease $T_c$ by 2--3 MeV, but not result in agreement with the EFT
prediction. An assumption that is made in the EFT prediction is that the
other LECs do not change appreciably as $d_3$ is taken to zero.  This may
not be accurate when extrapolating to the chiral limit from such large
values of $d_3$. Moreover, the extrapolation of \cite{Brandt:2013mba}
is also made from the same large input quark mass, and lattice results
may also shift considerably when lighter quarks are used. The fits of
the LECs for $N_f=2+1$ (see the next section) where the quark mass is
lighter shows very good agreement between the EFT prediction and the
direct lattice extractions, indicating that the higher quark masses here
are the cause of the mild disagreement. It would be interesting, when
future lattice computations of static pion properties become available,
to see how the LECs change as the pion mass is tuned on the lattice.

\subsection{Pion properties}

\bef
\includegraphics[scale=0.5]{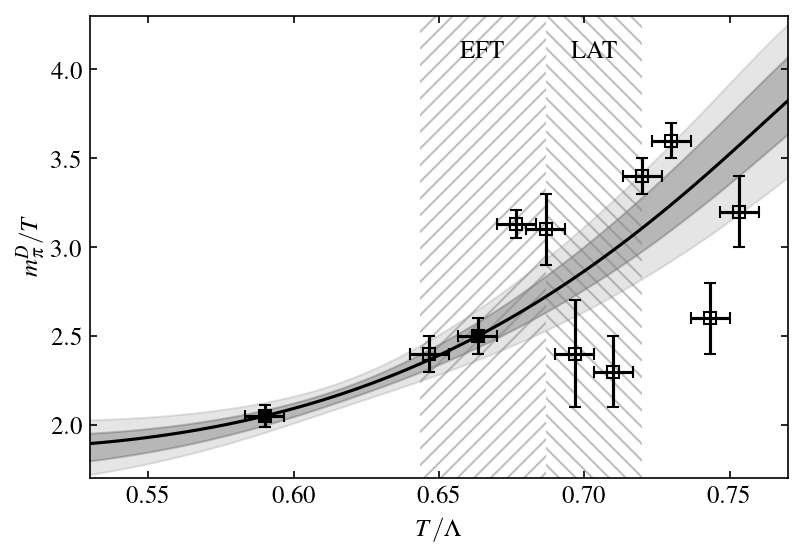}
\includegraphics[scale=0.5]{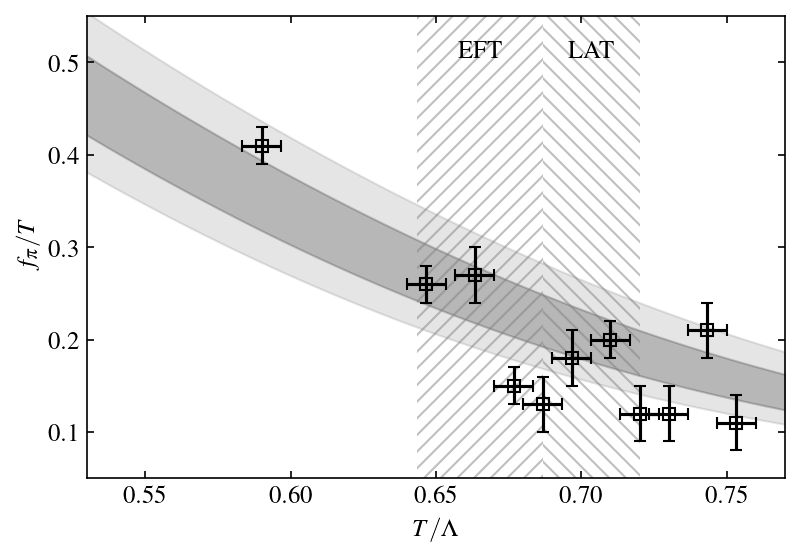}
\includegraphics[scale=0.5]{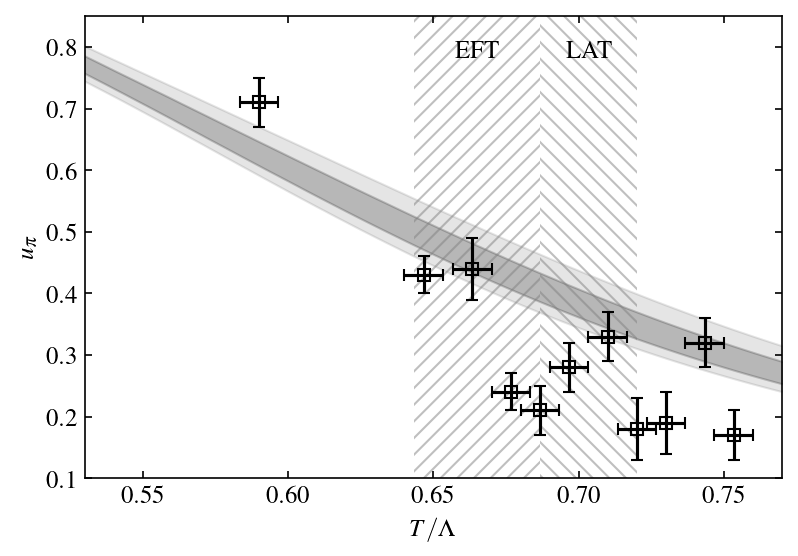}
\includegraphics[scale=0.5]{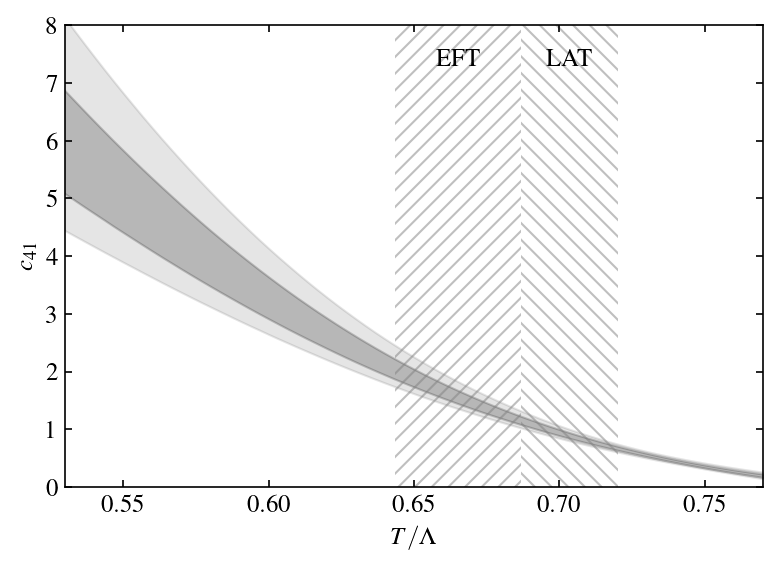}
\caption{Predictions for static pion properties from the EFT using the
LECs determined by the new scheme. The gray bands show the 68\% (darker
colour) and 95\% (lighter colour) CLs on the predictions. The vertical
bands are the predicted value of $\tco$ in the EFT, and the corresponding
lattice determination \cite{Brandt:2014qqa}. Two values of $m_\pi^D$ which
are inputs to the fits are shown as the filled points (the black line
shows the predictions from the best fit LECs). Clearly the prediction of
the temperature dependence of $m_\pi^D/T$, $f_\pi/T$ and $u_\pi$ are as
good as the data can support. The pion self-coupling $c_{41}$ has not
yet been measured on the lattice; we show the EFT prediction for it.}
\eef{pionpred}

It has been argued before that at a smooth cross over a description
of matter with hadron degrees of freedom may be useful even for
$T$ slightly larger than $\tco$. The failure of this picture will be
gradual. In \fgn{pionpred} one can see a remarkable ability of the
EFT to predict static pion properties at temperatures about 10--15\%
above $\tco$.  One may conclude from this that $d_3$, as shown in
\fgn{zerolecs}, is large enough for such a remarkable continuity of the
hadron description.

A diagnostic for this continuity is $u_\pi$. In the chiral limit it goes
to zero with a critical exponent \cite{Son:2002ci}, and the pressure and
various thermodynamic response functions have a singular behaviour. As
one sees in the lattice data, $u_\pi$ remains well above zero for this
simulation. The EFT prediction of $u_\pi$, shown in \fgn{pionpred} seems
to be a little too high.  Whether this is an artifact of our treatment
of the EFT in the Hartree-Fock approximation is an investigation that
we will return to in future.

In \fgn{pionpred} are also shown the EFT prediction of the pion self
coupling $c_{41}$. This has not been measured on the lattice since it
requires analysis of pion 4-point functions. However, the predictions are
reasonably accurate, and it seems to be worthwhile making the effort to
measure it on the lattice, since it is a completely independent test of
our quantitative understanding of the universal properties underlying
thermal pion physics and the phase diagram. Note that the predicted
values of $c_{41}$ are positive, whereas the relation of \eqn{coupling}
predicts a negative value. This implies that the terms in higher
powers of $m_\pi$ are important at these large values of $d_3$, and
these lattice simulations for $N_f=2$ are not very close to the chiral
limit. The mismatch between the lattice and EFT predictions of $T_c$
could be related. Nevertheless, the fact that so many predictions of
the EFT are in reasonable agreement with the measurements shows that
arguments based on chiral symmetry are a good guide to the essential
underlying physics.

\subsection{The pressure}

\bef
\includegraphics[scale=0.75]{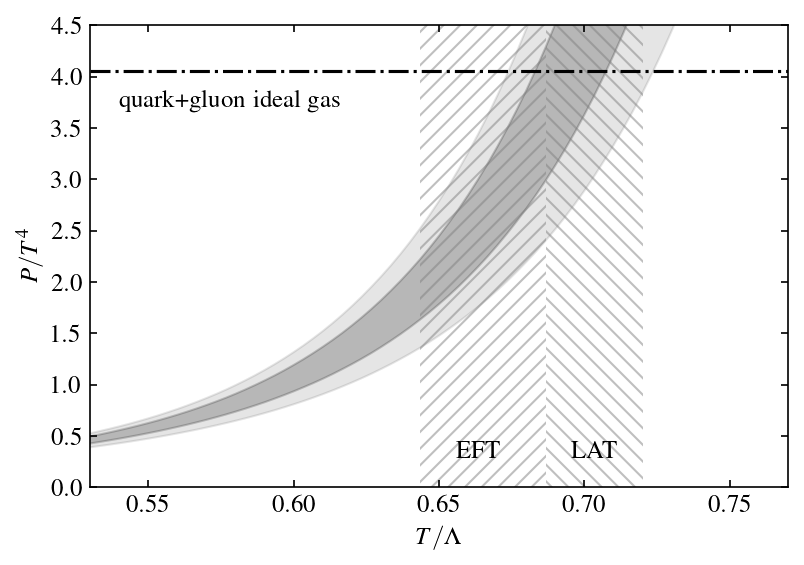}
\caption{The EFT predictions for the pressure are shown here. The 68\%
(darker colour) and 95\% (lighter colour) bands of prediction are shown.
The vertical bands show the predicted range of values for $\tco$ in
the EFT and on the lattice \cite{Brandt:2014qqa}. The rapid rise in the
predicted value of the pressure is noticeable.}
\eef{thermo}

The continuity of the hadron description encourages us to investigate
the thermodynamics of matter using this EFT. With the choice of $\Lambda$
well above $\tco$ this is even quantitatively possible. In \fgn{thermo}
we show the prediction for $P/T^4$ as a function of $T$. The rapid
rise in the pressure is a generic feature of such EFTs, and due to
the fact that the integral over spatial momenta of pion propagators
$1/(p_0^2+u_\pi^2p^2+m^2)$, can be converted to usual boson integrals
with the replacement $u_\pi p\to p$ (this is the equivalent of a similar
transformation used to obtain \eqn{enef}) . Through the volume element
$d^3p$ this then gives a factor of $1/u_\pi^3$ to the pressure. This
is one of the sources of the singularity in the pressure in the chiral
limit. In the Hartree-Fock approximation the critical exponent will have
the mean field value. An epsilon expansion would be needed to recover
the correct O(4) exponent \cite{Gupta:2015dra}.

The pressure reaches the limit of the ideal quark-gluon gas immediately
above $\tco$, indicating that there is a breakdown in the computation.
One cannot rule out the possibility that the range of applicability of
the EFT is different for each quantity. However, it will be discussed
in the section for $N_f=2+1$, where lattice measurements of $P/T^4$ are
available that there are more subtle problems which need to be resolved.

There is not only a rapid increase in $P/T^4$ as $T$ approaches $\tco$,
but also a rapid increase in its uncertainty. The 95\% CL band covers
almost a factor of two in $P/T^4$ near $\tco$. One notes two possible
origins for this error. One, of course, is that $u_\pi$ appears to a
high power in the expression for $P/T^4$ and therefore its uncertainty
is multiplied. The other is that $\tco$ is also uncertain and this may
produce part of the uncertainty in $P/T^4$.

\bef
\includegraphics[scale=0.5]{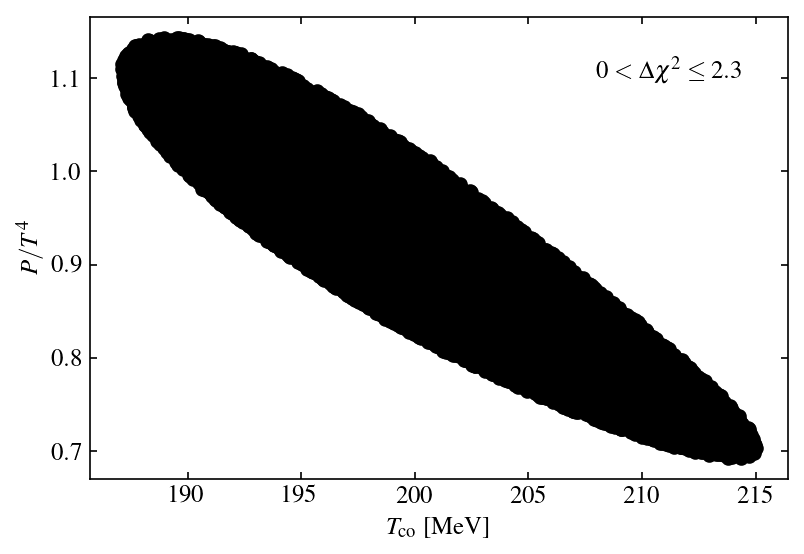}
\includegraphics[scale=0.5]{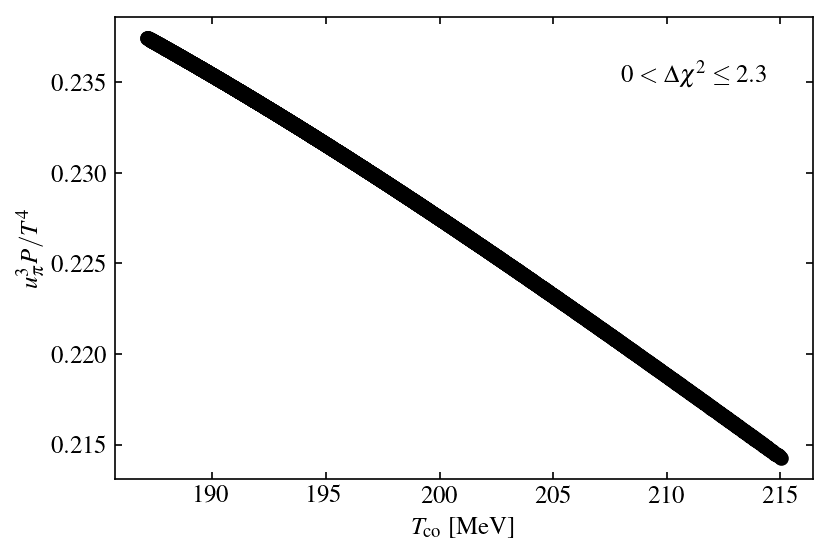}
\caption{The 68\% CL in $P/T^4$ at $T=177$ MeV is plotted against that
in $\tco$. Scaling $P$ by $u_\pi^3$ reduces the uncertainty band in
this direction, but the variation with $\tco$ is unchanged.}
\eef{pspread}

In \fgn{pspread} we show the 68\% CL on the joint distribution of $\tco$
and $P/T^4$ at a fixed $T$ below $\tco$. This is the distribution of
uncertainties propagated from the uncertainties in the fitted LECs. There
is clearly a correlation between them: as $\tco$ increases the pressure
decreases. But there is also a significant remnant uncertainty in
$P/T^4$.  The second panel shows the same region when plotted as the
joint distribution of $\tco$ and $u_\pi^3P/T^4$. This shows that a major
component of the uncertainty in $P/T^4$ is due to $u_\pi$. Once this
is removed, the dependence of $P/T^4$ on the particular combinations
of the LECs which give $\tco$ is much clearer. Measuring a variety of
pion properties on the lattice would let us test schemes for extraction
of the LECs which would best constrain the propagation of errors in
EFT predictions.

\goodbreak\section{Using lattice computations for $N_f=2+1$}\label{sec:four}

\bef
\includegraphics[scale=0.75]{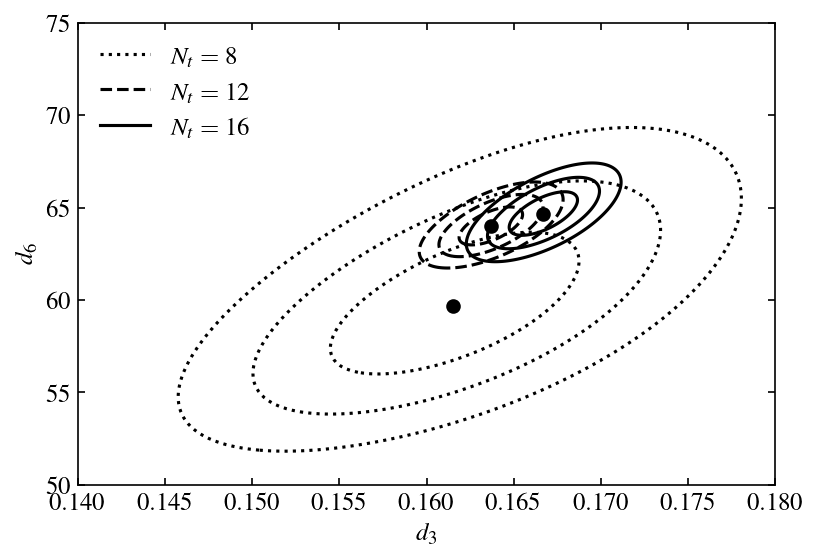}
\caption{The LECs fitted at $T=0$ for bare couplings which correspond to
nearly the same temperature in finite temperature lattices with Euclidean
time extent $N_t=8$, 12 and 16, each at the corresponding mass. The three
concentric ovals for each case are the 68\%, 95\% and 99\% CLs around
the best fit point at the center.  One sees that keeping $\Lambda=300$
MeV for the three different fits gives compatible results within 95\%
confidence intervals.
}
\eef{zerolecs3}

In this section we report the matching of the EFT to lattice data for
2+1 flavours. Our inputs are the lattice measurements of $m_\pi^D$
\cite{Bazavov:2019www}, and the predictions are tested by $\tco$
\cite{HotQCD:2018pds} and thermodynamics \cite{HotQCD:2014kol}. The lattice
measurements are presented for a physical value of the strange quark mass,
$m_s$, obtained by matching $m_\eta$ to its physical value. However, two
different light quark masses, $m_l$, are used. A choice of $m_l=m_s/27$
corresponds to the physical value of $m_\pi=140$ MeV, and the results
for $m_\pi^D$ \cite{Bazavov:2019www} and $\tco$ \cite{HotQCD:2018pds}
are given for this. However the thermodynamics \cite{HotQCD:2014kol} is
reported for a heavier light quark mass $m_l=m_s/20$. The pion properties
at $T=0$ have also been reported only for this heavier quark mass.  $\tco$
has also been reported for the heavier quark mass \cite{Bazavov:2011nk}.

\bet[b]
\begin{center}
\begin{tabular}{|c|c|c|c|c|c|c|c|}
\hline
$N_t$ & $T_{\rm Lat}^{\rm input}$ (MeV) 
	& $d_3$ & $d_4$  & $d_6$ 
	& $T_c$ (MeV) & $\tco$ (MeV) & $\tco^h$ (MeV)\\ 
\hline	
8 & 145, 156 & 0.120 & $ 1.30_{-0.05}^{+0.06} $ & $ 118_{-20}^{+25} $ & $ 141_{- 3, 5, 7}^{+ 3, 6, 9}$ & $ 166_{- 3, 5, 7}^{+ 3, 6, 9} $ & $ 172_{- 3, 5, 8}^{+ 3, 6, 9} $\\
& & 0.125 & $ 1.33_{-0.06}^{+0.06} $ & $ 130_{-22}^{+28} $ & $ 140_{- 3, 5, 7}^{+ 3, 6, 9} $ & $ 166_{- 3, 5, 8}^{+ 3, 6, 9} $ & $ 171_{- 3, 6, 8}^{+ 3, 6, 9} $ \\
\hline
12 & 145, 157 & 0.120 & $ 1.40_{-0.07}^{+0.09} $ & $ 169_{-32}^{+47} $ & $ 132_{- 3, 7, 9}^{+ 3, 6, 9} $ & $ 157_{- 3, 7, 9}^{+ 3, 7, 10} $ & $ 162_{- 3, 7, 10}^{+ 3, 7, 10} $\\
& & 0.125 & $ 1.44_{-0.07}^{+0.09} $ & $ 187_{-37}^{+54} $ & $ 131_{- 3, 7, 9}^{+ 3, 6, 9} $ & $ 157_{- 4, 7, 10}^{+ 3, 7, 10} $ & $ 162_{- 4, 7, 10}^{+ 3, 7, 10} $ \\
\hline
16 & 140, 152 & 0.120 & $ 1.37_{-0.11}^{+0.16} $ & $ 156_{-52}^{+95} $ & $ 130_{- 7, 15, 20}^{+ 7, 16, 27} $ & $ 155_{- 8, 16, 21}^{+ 8, 17, 28} $ & $ 160_{- 8, 16, 21}^{+ 8, 17, 28} $ \\
& & 0.125 & $ 1.41_{-0.12}^{+0.17} $ & $ 172_{-58}^{+109} $ & $ 129_{- 8, 16, 21}^{+ 8, 17, 27} $ & $ 155_{- 8, 17, 22}^{+ 8, 17, 28} $ & $ 160_{- 8, 17, 22}^{+ 8, 18, 29} $ \\
\hline
\end{tabular}
\end{center}
\caption{The table contains the LECs fitted to lattice data for $N_f=2+1$
for the set with $m_l=m_s/27$ using $\Lambda=300$ MeV. Also shown are
the EFT predictions for $\tco$ and $T_c$. For the last two the 68\%,
95\% and 99\% confidence limits are shown along with the bootstrap
median. The lattice determination is $\tco=156.5\pm1.5$ for this value
of the light quark mass \cite{HotQCD:2018pds}.  The ratio $\tco/T_c$
lies between 1.18 and 1.20 in all cases. By scaling $d_3$ in the ratio
of $m_l$, we also find $\tco^h$ corresponding to the heavier $m_l
= m_s/20$ light quarks.  For this case lattice measurements report
$\tco=155.9\pm8.0$ MeV \cite{Bazavov:2011nk}.}
\eet{fits}

We deal with this complication by assuming that the values of $d_3$ scale
in proportion to $m_l$, so we have for the heavier quark mass a value
$d_3^h$, and for the lighter physical quark mass we have $d_3=(20/27)
d_3^h$. We assume that the both $d_3$ and $d_3^h$ are light enough that
the other two LECs do not change when we go from one to the other. We
extract $d_3^h$ from the fits to $m_\pi$ and $f_\pi$ at $T=0$, and use
the scaled lighter value $d_3$ along with $m_\pi^D$ to extract $d_4$
and $d_6$. These can be used to predict $\tco$ as well as the other pion
properties for $T>0$. The extrapolation can be tested by using $d_3^h$
to compare the EFT prediction with lattice extractions of $\tco^h$. The
predictions for $P/T^4$ are also performed with $d_3^h$.

From $T=0$ hadron physics we know that at these values of $d_3$ chiral
power counting is accurate. Then the leading order chiral Lagrangian
for the thermal EFT is
\beq
  L_{\scriptscriptstyle LO} 
      = \frac12 c_2\Lambda^2(\pi^a\pi^a) 
      + \frac12(\partial_4\pi^a)(\partial_4\pi^a)
      + \frac12c_4 (\nabla_i\pi^a)(\nabla_i\pi^a),
\eeq{lpilo}
since $m_\pi^2=c_2\Lambda^2$ has the same scaling dimension as the two
derivatives in the kinetic terms. The term in $c_{41}$ in \eqn{lpi} is
one of several next-to leading order (NLO) terms which contribute to the
EFT. In view of this we treat both $m_\pi$ and $m_\pi^D=m_\pi/u_\pi$
without the Dyson-Schwinger resummation which was adopted for the
description of the lattice measurements for $N_f=2$ in \scn{three}.

\subsection{Extracting the LECs by matching lattice measurements}

\bef
\includegraphics[scale=0.75]{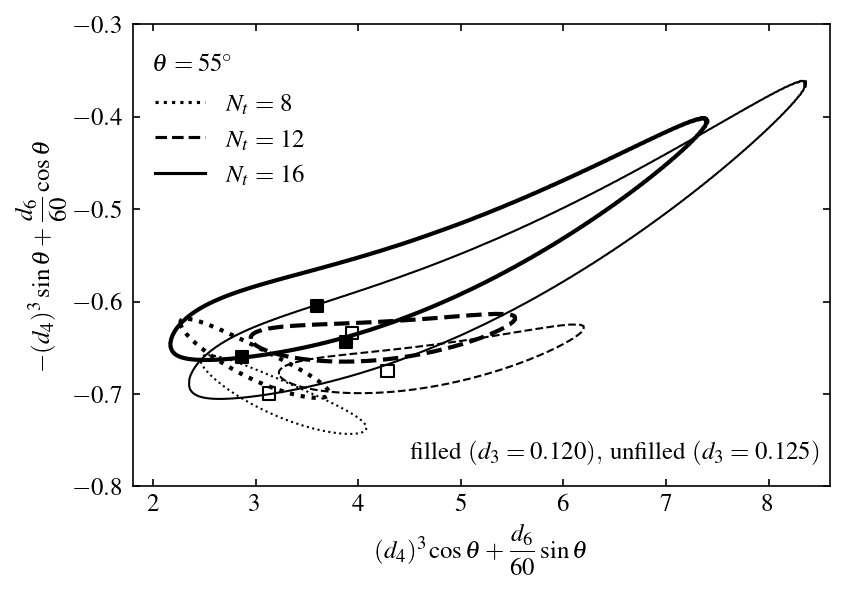}
\caption{The LECs $d_4$ and $d_6$ extracted by fits to lattice data for
$m_\pi^D$ for the two values of $d_3$ on three different $N_t$. The 
68\% confidence regions are marked (the thicker lines are for $d_3=0.120$
and thinner for $d_3=0.125$).}
\eef{fitcov}

As before, we will set the LEC $d_3$ using the $T=0$ lattice data which
are used to set the scale of the finite temperature computations. The
lattice has three bare parameters. Two are the light and strange quark
masses and these are used to tune the pion and other hadron masses. In
addition, the lattice has a bare coupling which can be traded for
the lattice spacing, $a$, which is the inverse of the UV cutoff of the
lattice computation. Ideally the continuum limit of lattice measurements
is taken by letting the lattice spacing go to zero (\ie, the lattice UV
cutoff go to infinity) while keeping physical quantities (such as the
$T$ and $m_\pi$) fixed.

When $\Lambda a\ll1$, \ie, the UV cutoff of the lattice is much larger
than the UV cutoff of the low energy EFT, then the process of taking
continuum limits can be made shorter, since lattices with different $a$
satisfying this condition are all equivalent as far as the low energy
effects are concerned. This has a practical consequence in fixing $d_3$,
as we show here.

With $\Lambda=300$ MeV as before, we extract $d_3$ and $d_6$ for the
$T=0$ EFT using lattice data with three different bare couplings fixed
so that $a=1/(T N_t)$ is roughly constant value of $T$ slightly below
$\tco$ for the three values of $N_t=8$, 12 and 16. The three $T=0$
simulations used correspond bare couplings which on the $N_t=8$ lattice
give $T=156\pm2$ MeV, on $N_t=12$ to $T=151.2\pm0.6$ MeV and for $N_t=16$
to $T=149.4\pm0.5$ MeV. The three temperatures are equal within 95\% CLs.
Note that $a$ changes by a factor of two, while $\Lambda a$ remains
significantly less than unity for all three simulations. The confidence
limits on the extracted LECs are shown in \fgn{zerolecs3}. The three
different sets of input data give LECs which are completely compatible
with each other, as shown.  Note, however, that the errors for the fit
at the lattice spacing corresponding to $N_t=8$ has much larger errors.
These are a consequence of the errors in the input measurements from the
lattice.

With the choice of $\Lambda=300$ MeV we use two representative values of
$d_3$, namely 0.120 and 0.125, in the extraction of the two remaining
LECs. We do this for one value of $N_t$ at a time.  For each $N_t$
we find that there are three measurements of $m_\pi^D$ reported in
\cite{Bazavov:2019www} for temperatures $T_1<T_2<T_3\le\tco$. For all three
$N_t=8$, 12 and 16, we chose to use the measurements at $T_1$ and $T_3$
(the values are given in the column marked $T_{\rm Lat}^{\rm input}$
in \tbn{fits}).  An estimate of the systematic errors is obtained by
changing the pair of input measurements in the fits. In all cases we
found that the extraction using $\{T_1,T_3\}$ lie between those using
the other two pairs, and this source of systematic uncertainty is a
little less than the 68\% confidence limits shown in \tbn{fits}.

Plotting the 68\% CLs in the $d_4$--$d_6$ plane shows a strong covariance
of the two LECs. Since the boundary of this region is not an ellipse, we
cannot use the usual second moment definition of the covariance matrix to
perform a principal components analysis. Instead it turns out to be useful
to rotate the axes by a numerically determined angle and plot the 68\% CLs
in this rotated frame, as shown in \fgn{fitcov}.

\bef[hb]
\includegraphics[scale=0.5]{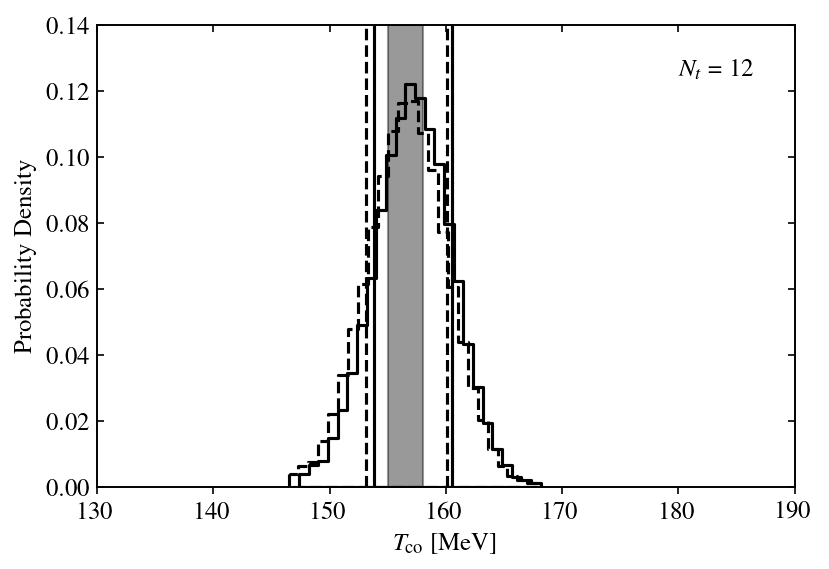}
\includegraphics[scale=0.5]{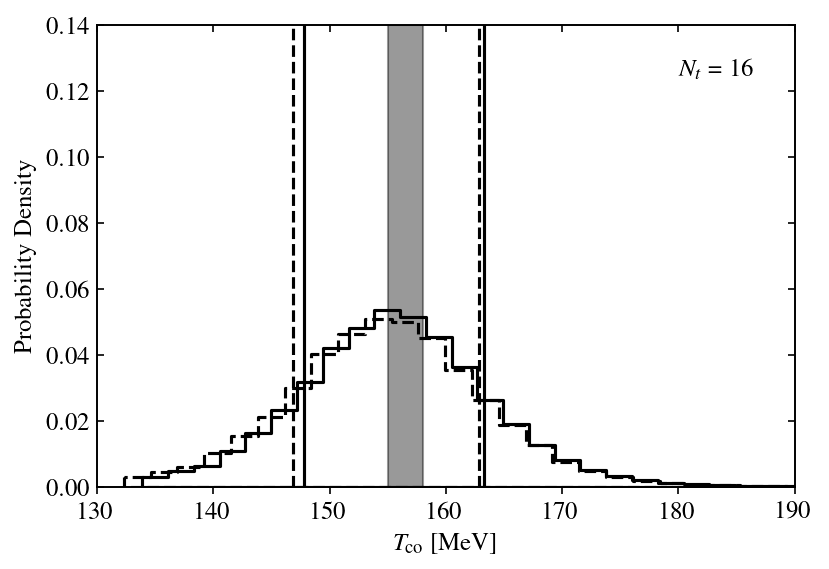}
\includegraphics[scale=0.5]{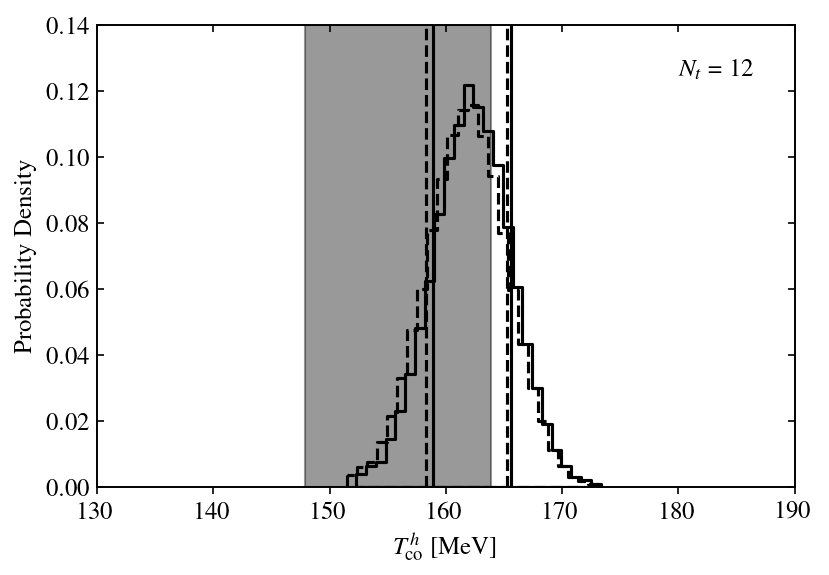}
\includegraphics[scale=0.5]{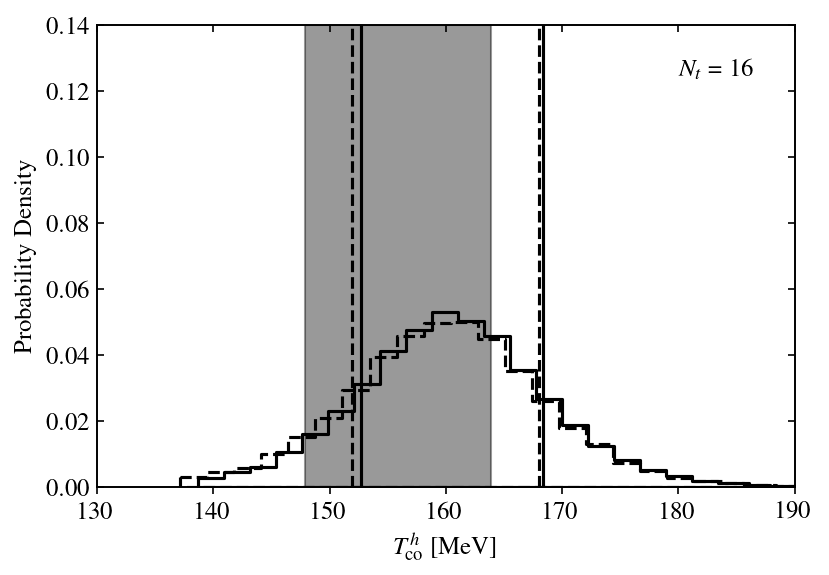}
\caption{Histograms for $\tco$ and $\tco^h$ obtained by sampling the
90\% CL of $d_4$ and $d_6$ for the two values of $d_3$ for $N_t=12$
and $N_t=16$. For each $N_t$ the full line gives the histogram and 68\%
confidence limit of $\tco$ for $d_3=0.120$ ($d_3=0.160$ for $\tco^h$)
and the broken line for $d_3=0.125$ ($d_3=0.169$ for $\tco^h$). For the
lighter quark the EFT predictions are in excellent agreement with the
lattice extraction \cite{HotQCD:2018pds}. The predictions for $\tco^h$
are also in agreement with the lattice extraction \cite{Bazavov:2011nk}.}
\eef{tco}

We notice several kinds of systematics here. First see that the 68\% CLs
of the fits for $d_3=0.120$ and 0.125 are compatible with each other.
Furthermore, the CLs for $N_t=12$ and 16 have very good overlap. Both
have some overlap with the contours for $N_t=8$, but the latter are
clearly to one side of the rest. The reason for this is not hard to find.
We find that the $T_{\rm lat}^{\rm input}$ for $N_t=8$ and 12 are close,
but the corresponding values of $m_\pi^D$ are quite different. 

This may be due to large lattice corrections in the pion correlation
function in going from $N_t=8$ to 12 which are not suppressed as the power
counting would seem to suggest. This is shown in \cite{Bazavov:2019www}
in terms of the taste-breaking of the pseudoscalar masses, where it is
shown that the RMS mass of the taste-partners is split by a much larger
amount for $N_t=8$ than for $N_t=12$ or 16. The large, but statistically
insignificant differences between the EFT predictions for $\tco$ and $T_c$
with $N_t=8$ and the rest, as shown in \tbn{fits}, then seems to be due
to a lattice artifact rather than a shortcoming of the EFT. In view of
this lattice uncertainty, in the remaining part of this section we only
show the EFT predictions using LECs fitted to $N_t=12$ and 16. These agree
for all the predictions we examined.

\subsection{The phase diagram from the EFT}

The distributions of the EFT predictions for $\tco$ with the LECs
obtained from lattice measurements of the two largest values of $N_t$
and with the two values of $d_3$ are shown in \fgn{tco}. Note the
excellent agreement between the EFT prediction of $\tco$ and the continuum
extrapolated lattice determination from \cite{HotQCD:2018pds}. Recall
that only pion properties have been used to determine the LECs, so this
is a good test of the EFT in two ways. The first is a quantitative test
of the underlying generalization of the universality argument which is
that pion properties are intimately connected with the phase diagram.
The second is the test that a single EFT with UV cutoff $\Lambda$
describes the long-distance behaviour of lattice computations with
different lattice UV cutoffs $1/a$, as long as pion properties do not
show unusual sensitivity to the dimensionless numbers $\Lambda a$.

\bef
\includegraphics[scale=0.5]{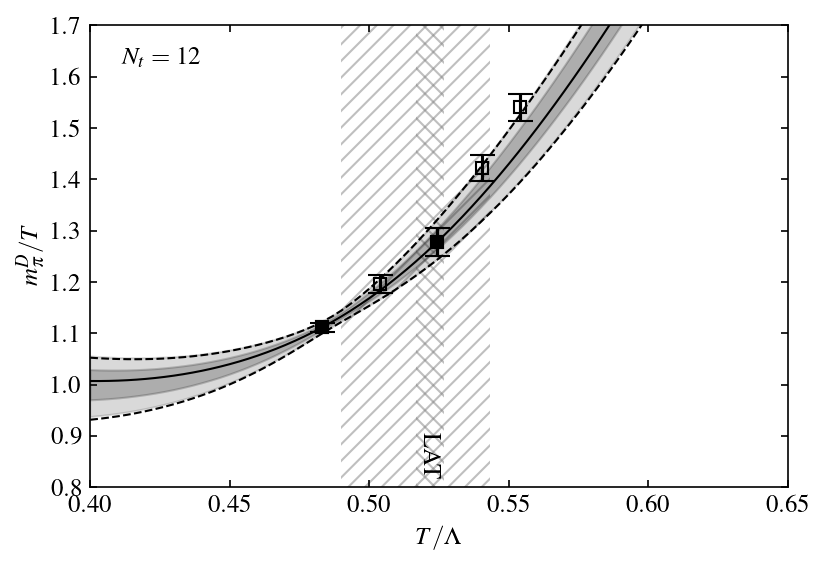}
\includegraphics[scale=0.5]{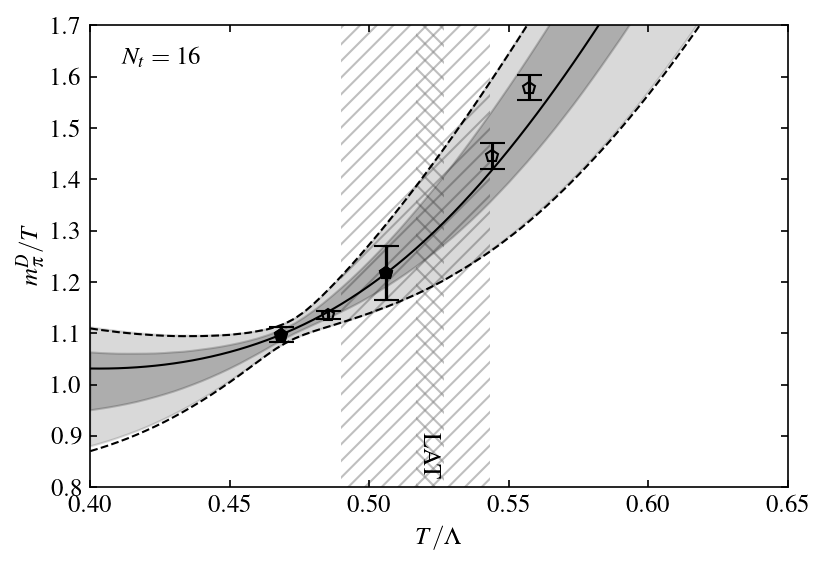}
\caption{EFT predictions for $m_\pi^D$ at finite lattice spacing for
$N_t=12$ and 16. The input data for extracting the LECs are shown with
filled symbols.  The prediction with the best fit LECs for $d_3=0.120$
are shown with the continuous line, and the 68\% (darker shade) and 95\%
(lighter shade) CL bands of the EFT predictions are also shown, along
with lattice measurements \cite{Bazavov:2019www}. The limits of the 95\%
CL band for $d_3=0.125$ are shown with dashed lines. The vertical bands
show the $\tco/\Lambda$ predicted by EFT and measured on the lattice.}
\eef{debyecutoff}

\bef
\includegraphics[scale=0.5]{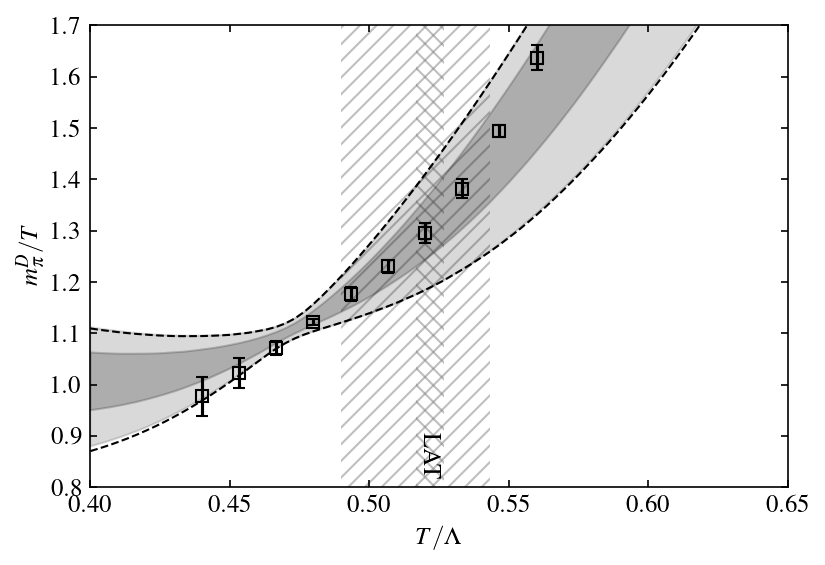}
\includegraphics[scale=0.5]{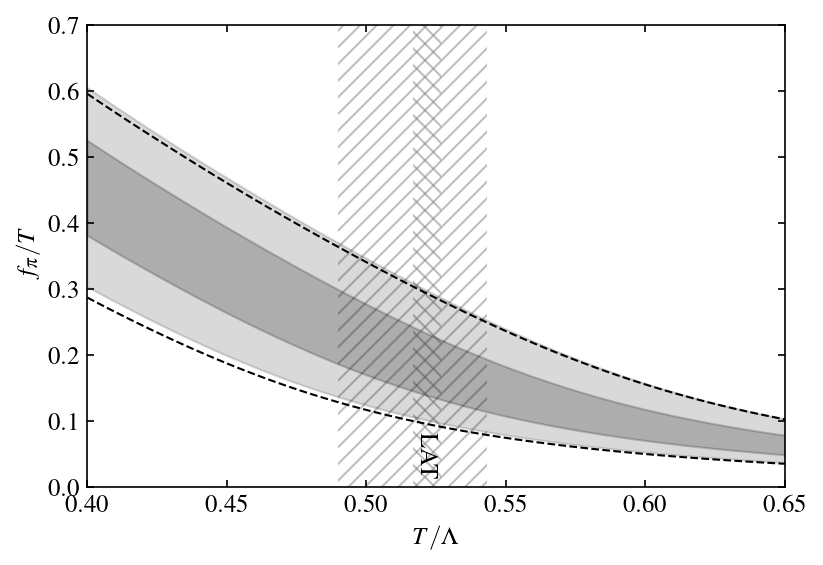}
\includegraphics[scale=0.5]{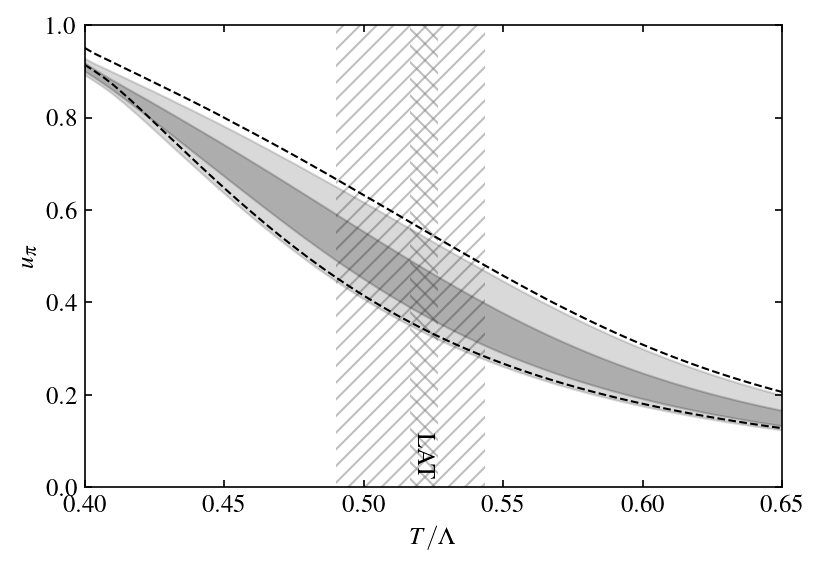}
\includegraphics[scale=0.5]{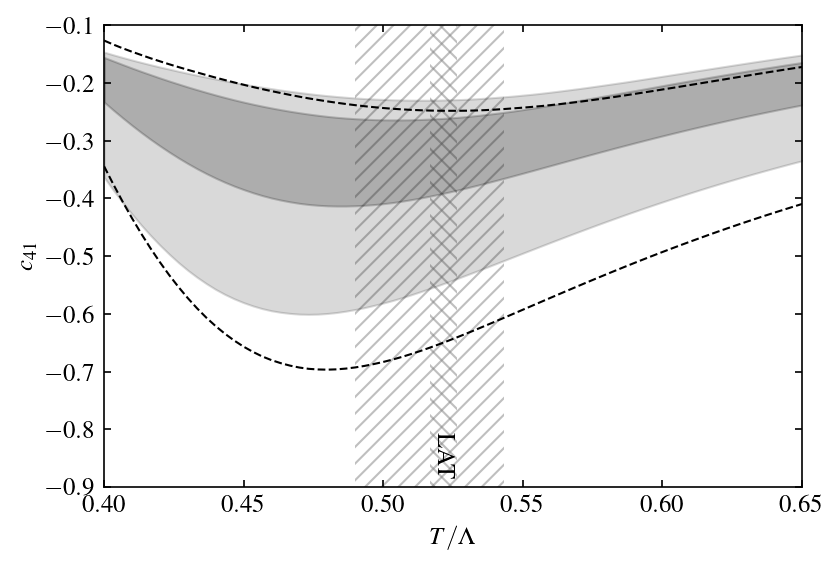}
\includegraphics[scale=0.5]{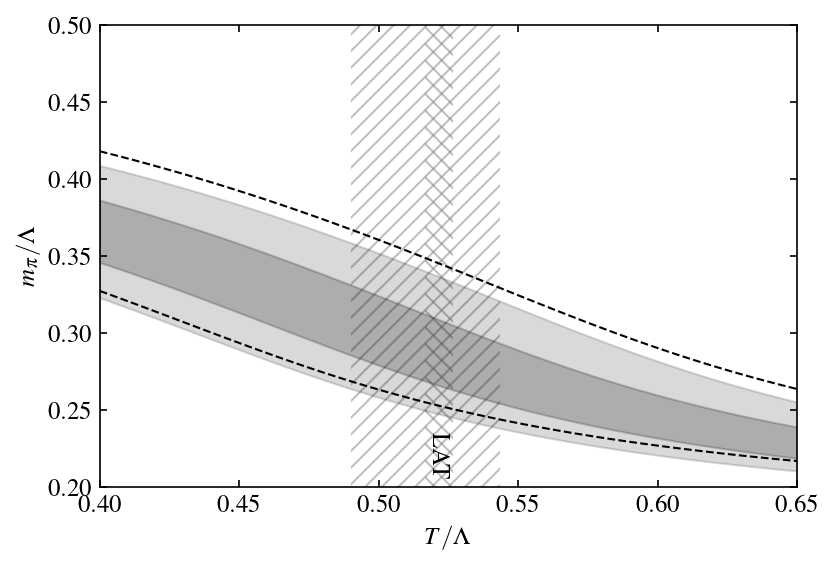}
\includegraphics[scale=0.5]{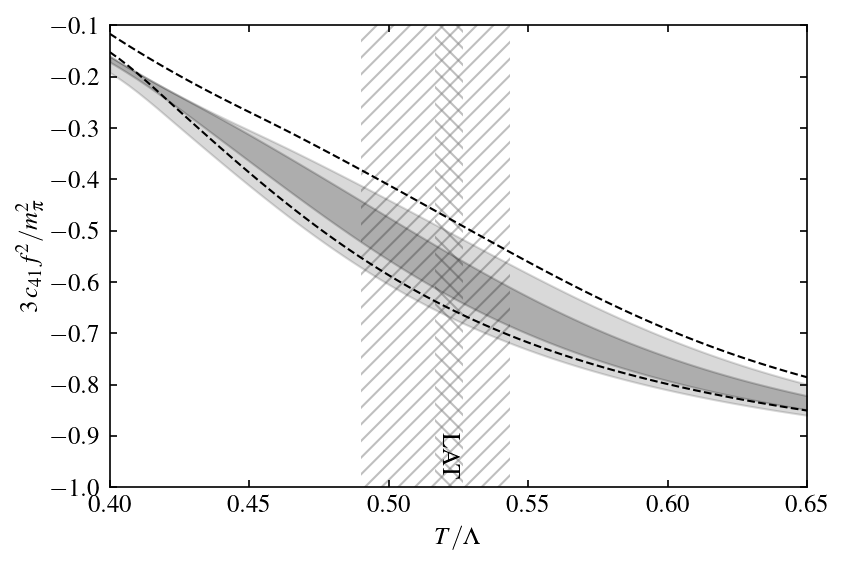}
\caption{EFT predictions for static pion properties are shown as the 68\%
(darker shade) and 95\% (lighter shade) CL bands when $d_3=0.120$. The
limits of the 95\% CL band for $d_3=0.125$ are shown with dashed lines.
The vertical bands show the $\tco/\Lambda$ predicted by EFT and measured
on the lattice \cite{HotQCD:2018pds} (the latter is entirely contained
within the former). The continuum extrapolated values of $m_\pi^D$
reported in \cite{Bazavov:2019www} are shown. Note that at the lowest
temperatures this continuum extrapolation used lattice spacing with $N_t=8$
or coarser. The bottom two panels show that the pion mass is large enough
that NLO terms in the thermal chiral EFT may be numerically important.}
\eef{pion21}

In \fgn{tco} we also show the histogram for $\tco^h$ predicted by the EFT
when $d_3$ is scaled in the ratio $27/20$, which is the ratio of the light
quark masses for the extraction of the remaining LECs, and the lattice
extraction of $\tco$ in \cite{Bazavov:2011nk}. Again in this case, there
is agreement between the EFT prediction of $\tco^h$ and the lattice
determination. In future if the screening masses for the heavier quark
mass are published then a direct fit can be used to further check this
result. The agreement of the EFT and lattice determination of $\tco$ in
these two cases leads us to believe that the mild disagreement for $N_f=2$
is the result of the light quark mass being significantly higher there.

The EFT also gives the prediction for $T_c$ extrapolated to zero
quark mass, $T_c=131^{+7}_{-6}$ MeV. This can be compared to the value
$T_c=132^{+3}_{-6}$ MeV quoted in \cite{HotQCD:2019xnw} obtained by an
extrapolation of the lattice data to the continuum and then to chiral
limit using O(4) scaling.  Recall that the prediction for the
curvature coefficient $\kappa_2$ was given in \eqn{eftpred} and that
$\widetilde\kappa_4=0$. Both these results are in good agreement with
$N_f=2+1$ lattice measurements collected in \tbn{curvs}. This completes
the EFT predictions for the phase diagram.

\subsection{Pions}

With two lattice measurements as input, the primary predictions of
the EFT are the values of the four LECs for the pion Lagrangian in
\eqn{lpi}, and the derived quantities $m_\pi^D=m_\pi/u_\pi$, the pion's
Debye screening mass, and $f_\pi=f u_\pi$, the finite temperature pion
decay constant. For $N_f=2+1$, of these four quantities only $m_\pi^D$
has been measured. We show the prediction for this Debye screening
mass against the measurements on the lattice for $N_t=12$ and 16 in
\fgn{debyecutoff}. Notice that the difference between the fit uncertainties
for the $N_t=12$ and 16 lattices are simply propagated from the substantially
larger error in one of the input data for the $N_t=16$ lattices. As one
can see in \fgn{debyecutoff} the only effect is in the larger error bars for
predictions.

Even this close to the chiral limit, there is good agreement with the
screening masses above $T_c$. At a crossover one does not expect abrupt
changes in the description of matter, but it is nevertheless surprising
to see the quantitative agreement between the EFT and lattice data
at temperatures more than 5\% above $\tco$. This could well be due
to using the Lagrangian in \eqn{lpilo}. Adding NLO terms could modify
this behaviour.

Whether the $N_t=12$ lattice measurements are used as input or that from
$N_t=16$, the EFT predictions of other pion properties do not change. In
\fgn{pion21} we show these predictions for $m_\pi^D$, $f_\pi$, $u_\pi$
and $c_{41}$ using the LECs extracted by matching to $N_t=16$. Also shown
are continuum extrapolations of lattice measurements of $m_\pi^D$ as given
in \cite{HotQCD:2018pds}.  The minor, and statistically insignificant,
mismatch at the lowest $T$ then is a little surprising at first. However,
we find that the continuum extrapolation at this temperature is obtained
from coarse lattice with $N_t=8$ at best.  Large lattice artifacts for
these coarser lattice spacings have been discussed already.

In \fgn{pion21} we also show the pole mass, $m_\pi=\sqrt{c_2}\Lambda$.
Note the falling trend as it approaches $\tco$. In physical units the
EFT predicts $m_\pi\simeq100$ MeV at $\tco$ (the screening mass is
about twice as large). When the temperature is 15\% lower,
the pole mass climbs to about 110 MeV (and the screening mass is about
135 MeV). The pole masses are significantly less than the $T=0$ value
of the mass, but they are nevertheless of order $T$. In the chiral limit
the pole mass would vanish.

The results for $c_{41}$ shown in \fgn{pion21} were computed using
all terms in the expansion in $d_3$, and not just the leading two
terms shown in \eqn{coupling}. Close enough to the chiral limit, \ie,
when $d_3$ is small, \eqn{coupling} implies that $3c_{41}f^2/m_\pi^2$
should be close to $-1$. In the final panel of \fgn{pion21} we show
that numerically this is far from exact. In fact $c_{41}$ is expanded in
powers of the dimensionless ratio $m_\pi/f_a = u_\pi^2 m_\pi^D/f_\pi$,
which turns out to be close to unity at $\tco$.  So the higher order
terms in the series for $c_{41}$ are not parametrically suppressed. Much
lower values of the pion mass at $T=0$ would be required for the leading
term to be numerically accurate at all temperatures.

One knows from current algebra phenomenology at $T=0$ that the assumption
of broken chiral symmetry leads to some strikingly good results. However,
in other domains these predictions were not quantitatively reliable. Today
we understand that higher order terms in chiral perturbation theory are
needed to reach the same level of accuracy in other predictions. The
situation seems to be similar at finite temperature. A key question
seems to be how small a ratio like $m_\pi/f$ needs to be.

\subsection{The pressure}

\bef
\includegraphics[scale=0.75]{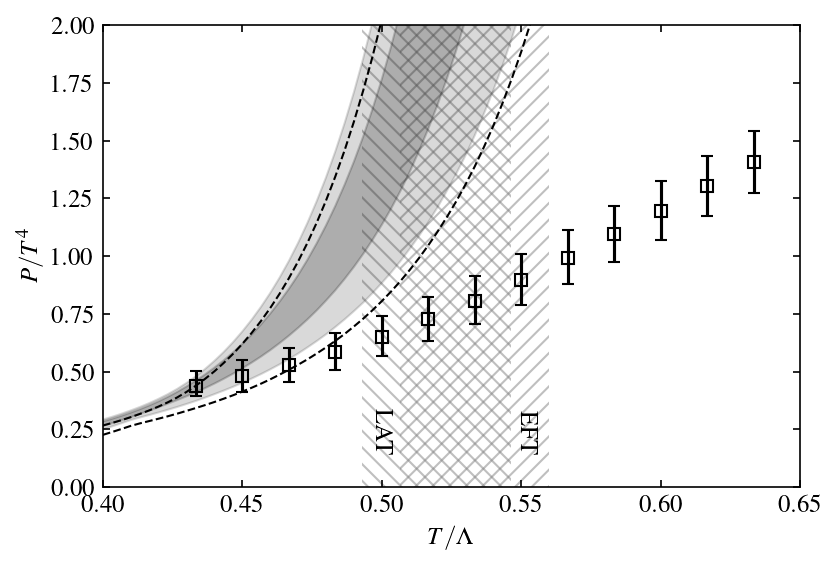}
\caption{EFT predictions for $P/T^4$ are shown as the 68\% (darker shade)
and 95\% (lighter shade) CL bands when $d_3^h=0.160$. The limits of the 95\%
CL band for $d_3^h=0.169$ are shown with dashed lines.  The vertical bands
show the $\tco/\Lambda$ predicted by EFT and measured on the lattice
\cite{Bazavov:2011nk}. The continuum extrapolated values of $P/T^4$
reported in \cite{HotQCD:2014kol} are shown.}
\eef{pressure21}

The pressure of strongly interacting matter is another prediction from the
EFT. The results from lattice measurements are plotted in \fgn{pressure21}
along with the prediction from $L_{\scriptscriptstyle LO}$ of \eqn{lpilo}.
Since this is a quadratic Lagrangian, the result is the ideal gas
pressure apart from the factor of $1/u_\pi^3$ which has been discussed
previously. The EFT gives a quantitatively reliable prediction of $P/T^4$
for $T/\Lambda<0.5$. 

Thereafter, the rapid rise in the prediction of $P/T^4$ visible in
\fgn{pressure21} is mainly due to the drop in $u_\pi$ as one approaches
$\tco$. The effect of the drop in $m_\pi$ with $T$ is subleading. In
\scn{two} we argued that at least a 2-loop resummation of the Dyson
Schwinger equation for the Lagrangian in \eqn{lpi} is needed to change
$u_\pi$ from its tree level value.

The formal argument remains valid even when the term in $c_{41}$ has
to be included with chiral power counting. However, in that case one
has to account for all the other NLO terms in the thermal chiral EFT.
In the $T=0$ chiral perturbation theory the unitarized resummation
of all these terms gives rise to the resonance spectrum of mesons
\cite{GomezNicola:2001as}. In this sense it seems that a higher order
computation of the pressure in the EFT could be formally equivalent
to a computation in an interacting resonance gas described by a finite
temperature chiral EFT. This is an extension of the chiral EFT approach
that we leave to the future.

\goodbreak\section{Conclusions}\label{sec:five}

\bef[tb]
\includegraphics[scale=0.75]{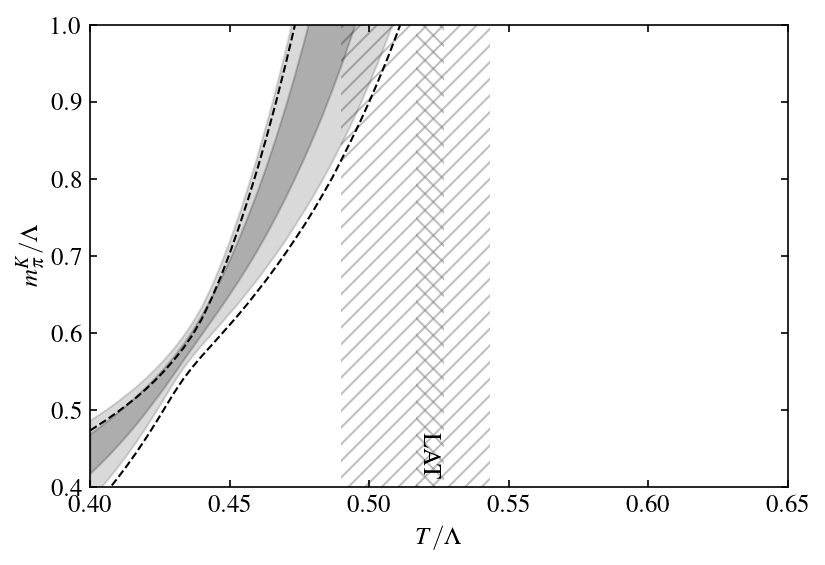}
\caption{The pion kinetic mass, $m_\pi^K$, in the EFT taken at leading
order in chiral power counting for $N_f=2+1$ QCD with realistic pion and
Kaon masses.}
\eef{mkin}

We described here a thermal EFT for $N_f=2+1$ flavours of
interacting quarks which we treated in the Hartree-Fock approximation
(see \scn{two}.A).  We then obtained (in \scn{two}.B) an EFT for
the pseudo-Goldstone bosons, which are the small fluctuations around
the solution of the resulting gap equation. This appears in the form
of a thermal chiral perturbation theory (T$\chi$PT) with an octet
of pseudoscalar mesons which can be matched to lattice computations.
We also argued that it can be reduced further to a T$\chi$PT involving
only pions.  UV insensitivity of low-energy EFTs then allows us to treat
the pion theory as descending from an effective $N_f=2$ quark theory
whose LECs contain the information of the effects of the strange quark
(see \scn{two}.C). From this we found an expression for the shape of the
phase boundary which has an interesting large $N_c$ limit. With increasing
$N_c$ the phase boundary first approaches an elliptical shape, which then
flattens out, with $T_c$ becoming independent of $\mu_\B$.  The lattice
measurements \cite{HotQCD:2018pds, Borsanyi:2020fev} of $\kappa_4$
can be understood in the context of the EFT from large $N_c$ counting.

Different schemes for extracting the LECs from lattice inputs give
essentially the same results. One sees this by comparing the results given
for $N_f=2$ in \scn{three} with the fits given in \cite{Gupta:2017gbs},
where the LECs are extracted in different ways. The comparison shows
that different ways of extracting the LECs can give rise to different
uncertainties in predictions.

Our main new results are the description of lattice measurements for
$N_f=2+1$. These are given in \scn{four}, and use the method of extracting
the LECs which was tested in \scn{three}.  Our first interesting
observation is that the thermal chiral EFT gives essentially the same
predictions whether input data is taken from lattices with $N_t=12$ or
16. This is understandable since the UV cutoff of the EFT, $\Lambda$,
is much smaller than that of either lattice. A subtlety with $N_t=8$
is discussed in \scn{four}.

With static pion properties as input, the EFT is able to predict results
for the phase diagram of QCD which are in good agreement with the direct
lattice measurements (see \fgn{tco}).  It is interesting to see that
the prediction of both $\tco$ and its chiral extrapolation, $T_c$, agree
extremely well with the lattice extractions.  This implies that the EFT
is in the region where $d_3$ is small enough for chiral symmetry to be
quantitatively useful already at the leading order of power counting in
the EFT.

The EFT also predicts other static properties of the pion which are not
currently available through lattice measurements. Among these we count
the pion decay constant at finite temperature, $f_\pi$, the pion thermal
``velocity'', $u_\pi$, the pole mass, $m_\pi$, and the pion self coupling,
$c_{41}$. These quantities are defined in \scn{two} and the predictions
are shown in \fgn{pion21}.

On the other hand, $m_\pi/f$ is of order unity, as a result of which
the leading chiral expression for $c_{41}$ in \eqn{coupling} is not
numerically accurate.  Similar behaviour has been seen in hadron
phenomenology at $T=0$ where some quantities are well described by LO
chiral perturbation theory, but others require at least an NLO treatment.

The attempt to describe the pressure of strongly interacting matter is
shown in \fgn{pressure21}. We note that the EFT is able to quantitatively
capture the behaviour of $P/T^4$ for $T/\Lambda<0.5$. Beyond this the
EFT prediction rises much faster than that measured on the lattice. We
argue that this is an NLO effect. This computation is substantial,
and outside the scope of this paper. So it is left for the future.

A domain where lattice computations are unreliable is in the analytic
continuation to real time. However, the analytic continuation of the
EFT is straightforward \cite{Gupta:2020zqo}. For example, the dispersion
relation of the pion in \eqn{lpilo} is
\beq
  E_p = \sqrt{m_\pi^2+u_\pi^2p^2} \simeq m_\pi + \frac{p^2}{2m_\pi/u_\pi^2}
   + \cdots
\eeq{kinetic}
at low momenta $p\ll\Lambda$. This means that the kinetic energy involves
a kinetic mass $m_\pi^K=m_\pi/u_\pi^2$.  We show our prediction for
this quantity for $N_f=2+1$ with realistic $T=0$ masses of the pion
and Kaon in \fgn{mkin}.  Note the very rapid rise in the kinetic mass
as $T$ increases. This rise may be moderated when NLO power counting
terms are included in the EFT. Although the numbers may change,
the fact that $u_\pi$ would fall close to $T_c$ means that $m_\pi^K$
is bound to increase.  It is interesting to note that this means that
with increasing $T$ the kinetic energy added by an increase in momentum
decreases. Reactions which were possible at low temperature might be
blocked due to this reason at finite temperature. The unexpected
coexistence in chiral symmetry restored matter of a slow rise of the
screening mass, implying the presence of pion collective excitations,
and a rapid rise of the kinetic mass, implying its decoupling from the
dynamics, points to a complex picture of strongly interacting matter
across the crossover.

In summary then, based on the chiral symmetry of quarks we wrote a finite
temperature EFT which took input from a small number of static pion
properties computed for $N_f=2+1$ QCD in equilibrium with realistic pion
and Kaon masses (at $T=0$). This gave predictions of the QCD phase diagram
with a leading order computation, which were in excellent agreement with
lattice measurements. The EFT also made predictions for other static
pion properties which can be tested in future lattice computations.
We noted that the errors of the EFT predictions are due to propagation
of errors from the inputs. Therefore, improved measurements of $m_\pi^D$
can substantially improve the test of the EFT predictions. We noted that
some quantities like the pressure of strongly interacting matter and the
real time quantity called the kinetic mass, defined in \eqn{kinetic},
may require an NLO computation in the EFT. This is a future research
direction.

\appendix
\goodbreak\section{Curvature coefficients}\label{apx:one}

The change in $T_c$ with the baryon chemical potential, $\mu_\B$, has
been used to define the curvature coefficients
\beq
  T_c(\mu_\B) = T_c\left[1 - \kappa_2 \left(\frac{\mu_\B}{T_c}\right)^2
	- \kappa_4 \left(\frac{\mu_\B}{T_c}\right)^4 + \cdots\right]
\eeq{curvs}
in agreement with the notation of \cite{HotQCD:2018pds,
Borsanyi:2020fev}. In terms of derivatives we have
\beq
  \kappa_2 = - T_c \left.\left(\frac d{d\mu_\B^2}\right)\,T_c(\mu_\B)
     \right|_{\mu_\B=0}
  \qquad{\rm and}\qquad
  \kappa_4 = - \frac12T_c^3 \left.\left(\frac d{d\mu_\B^2}\right)^2\,T_c(\mu_\B)
     \right|_{\mu_\B=0}
\eeq{derivs}
so that the curvature coefficients are explicitly dimensionless.
Note that the derivatives are taken with respect to a variable $\mu_\B^2$.

Comparing this with the chiral critical ellipse, which is the phase
diagram of the NJL-like models, one can quantify the departure from
ellipticity in terms of the parameter
\beq
  \widetilde\kappa_4 = \kappa_4-\frac12\kappa_2^2.
\eeq{tilded}
Lattice measurements of $\kappa_2$ began to converge to a common value
following the work of \cite{Cea:2014xva,Bonati:2014rfa}. In recent years
the value of $\kappa_4$ has also been reported. In \tbn{curvs}
we collect all the recent measurements that we are aware of.

\goodbreak\section{Changing $\Lambda$}\label{apx:two}

We noted that the UV cutoff $\Lambda$ used to define the EFT can be chosen
to be anywhere between the pion and Kaon masses. In this sense it is a
pseudo-parameter: a different choice of $\Lambda$ would change the LECs
but not the predictions. This is the meaning of a renormalization group
(RG) flow in an EFT.

\bet
\begin{center}
\begin{tabular}{|c|c|c|c|c|c|c|c|}
\hline
$N_t$ & $T_{\rm Lat}^{\rm input}$ (MeV) 
	& $d_3$ & $d_4$  & $d_6$ 
	& $T_c$ (MeV) & $\tco$ (MeV) & $\tco^h$ (MeV)\\ 
\hline	
12 & 145, 157 & 
	$0.09$ & $ 1.51_{-0.09}^{+0.12} $ & $ 520_{-119}^{+194} $ & 
	$ 125_{- 4, 8, 11}^{+ 4, 7, 11} $ & 
	$ 153_{- 4, 9, 12}^{+ 4, 8, 12} $ &
	$ 159_{- 4, 9, 12}^{+ 4, 8, 12} $ \\
16 & 140, 152 &
	$0.09$ & $ 1.45_{-0.14}^{+0.22} $ & $ 455_{-169}^{+363} $ &
	$ 123_{- 9, 18, 21}^{+ 9, 18, 27} $ &
	$ 151_{- 10, 19, 22}^{+ 9, 19, 29} $ &
	$ 157_{- 10, 19, 23}^{+ 9, 19, 29} $ \\
\hline
\end{tabular}
\end{center}
\caption{The table contains the LECs and $T_c$, $\tco$ and $\tco^h$
with for $N_f=2+1$ for UV cutoff $\Lambda=450$ MeV. The input data from
lattice measurements is exactly as for $\Lambda=300$ MeV.}

\eet{rgflow}

We demonstrate this in $2+1$ QCD with the alternate choice of
$\Lambda=450$ MeV.  The fit to the same $T=0$ data for pions used in the
main text changes the best fit values of $d_3^h$. Using the scaled $d_3$
and the same inputs for $T>0$ lattice data as before, we find that the
best fit LECs change substantially. However, as can be seen by comparing
the results in \tbn{fits} and \tbn{rgflow}, the predictions for $\tco$
and $\tco^h$ are unchanged within errors.  

There is a downward movement in the extrapolation of $T_c$ in the limit
of massless quarks, but this is also within the 95\% CL of the lattice
fits. In any case, such minor differences in predictions with two values
of the cutoff are expected when the EFT is treated approximately. Even
for perturbative QCD, changing the renormalization scheme changes the
results of finite order perturbative predictions \cite{Stevenson:1981vj};
only all orders predictions are expected to be precisely unchanged.

\goodbreak\section{Loop integrals}\label{apx:four}

For loop integrals in thermal EFT we follow the notation and procedure
of \cite{Gupta:2017gbs}. Since we deal only with one-loop contributions,
there is only a single loop momentum to integrate over, the 4-momentum
$p=(p_4,{\bf p})$ in the following.  Integrals over 4-momenta mean a
sum over Matsubara modes and integral over three momenta. We will need
the three basis integrals
\beqa
\nonumber
 J_0^{ab} &=& \frac{N_c}{\Lambda^2}\int\frac{d^4p}{(2\pi)^4}
     \frac{m_am_b}{(p_a^2+m_a^2)(p_b^2+m_b^2)},\\
\nonumber
 J_1^{ab} &=& \frac{N_c}{\Lambda^2}\int\frac{d^4p}{(2\pi)^4}
     \frac{(p_a)_4(p_b)_4}{(p_a^2+m_a^2)(p_b^2+m_b^2)},\\
 J_2^{ab} &=& \frac{N_c}{\Lambda^2}\int\frac{d^4p}{(2\pi)^4}
     \frac{|{\bf p}_a|\,|{\bf p}_b|}{(p_a^2+m_a^2)(p_b^2+m_b^2)}
\eeqa{basisints} 
where the two quarks $a$ and $b$ have momenta $p_a$ and $p_b$, and can
be either light or strange flavours, with $m_a$ and $m_b$ taking the
appropriate values. Furthermore, we have zero external momentum at the
vertices, so we can take $|{\bf p}_a|=|{\bf p}_b|$. The integrals have
been rendered dimensionless using powers of $\Lambda$. The overall factors
of the number of colours, $N_c$, and the dimension of the Dirac spinor,
$N_s$, come from the trace over all components of the quarks. The trace
over flavours is complicated because of the splitting of strange and light
flavours and the factors coming from them will be written explicitly when
the LECs are written.

There are possible UV divergences in the vacuum parts of the loop
integrals, and they are treated in dimensional regularization (see
\cite{Gupta:2017gbs}). There are no UV divergences in the thermal parts
of the integrals since they are regulated by the Fermi distribution
which arises from the Matsubara sum. It is also readily checked that
IR divergences do not arise in any of the three integrals. The zero
temperature pieces of the integrals have powers of $m$ multiplying any
$\log m$ that appears. So all of these integrals are regular in the
chiral limit.

The integrals that are needed can be written in terms of these basis
integrals. For examples, in order to write the scale factor and LECs
we need the one-loop integrals
\beq
 \I_1^{ab} = J_0^{ab} + J_1^{ab} + J_2^{ab}, \quad
 \I_2^{ab} = J_0^{ab} - J_1^{ab} - J_2^{ab},\quad
 \I_3^{ab} = -J_0^{ab} + J_1^{ab} + \frac13J_2^{ab},\quad
 \I_4^{ab} = -J_0^{ab} - J_1^{ab} + J_2^{ab}.
\eeq{twoints}
For $f_a$, $c_2^a$, $c_4^a$ and $c_{41}^a$ both quarks are light in
all the integrals, for the corresponding LECs for Kaons, one of the
quarks is strange, and so on. The notation of \cite{Gupta:2017gbs}
was $\I$ instead of $\I_1^{\ell\ell}$, $\I_{ii}$ for $I_3^{\ell\ell}$
and $\I_{44}$ instead of $I_4^{\ell\ell}$.

\goodbreak

\end{document}